\documentclass[numbook, envcountsect, envcountsame, envcountreset, runningheads, smallextended]{svjour3}

                                            \usepackage[numbers]{natbib} 

\usepackage{graphicx}
\usepackage{amsfonts}
\usepackage{amssymb}
\usepackage{amsmath}
\usepackage{enumerate}
\usepackage{eurosym}
\usepackage{url}

\newcommand{\R}{\mathbb{R}}
\newcommand{\E}{\mathbb{E}}
\newcommand{\N}{\mathbb{N}}
\newcommand{\Prob}{\mathbb{P}}
\newcommand{\Q}{\mathbb{Q}}

\newcommand{\1}{\mathbf{1}}
\newcommand{\CF}{\mathcal{F}}

\newcommand{\Qd}{\Q^{\$}}
\newcommand{\eu}{\text{\euro}}
\newcommand{\Qe}{\Q^{\eu}}
\newcommand{\dd}{\mathrm{d}}
\newcommand{\cadlag}{c\`adl\`ag}

\newcommand{\tInd}{_{t \in [0,T]}}

\allowdisplaybreaks
\smartqed

\title{On the hedging of options on exploding exchange rates\footnote{The views represented herein are the authors' own views and do not necessarily represent the views of Morgan Stanley or its affiliates and are not a product of Morgan Stanley research.}}
\titlerunning{On the hedging of options on exploding exchange rates} 
\author{Peter Carr\and
    Travis Fisher\and
    Johannes Ruf}
\institute{Peter Carr 
\at New York University, Courant Institute\\\email{pcarr@nyc.rr.com}
\and
Travis Fisher
\at \email{traviswfisher@gmail.com}
\and
Johannes Ruf
\at University of Oxford, Oxford-Man Institute of Quantitative
    Finance and Mathematical Institute \\\email{johannes.ruf@oxford-man.ox.ac.uk}
}

\begin{document}
 \maketitle

\begin{abstract}
\noindent
We study a novel pricing operator for complete, local martingale models.   The new pricing operator guarantees
put-call parity to hold for model prices and
the value of a forward contract to match the buy-and-hold strategy, even if the underlying follows strict local martingale dynamics.
More precisely, we discuss a change of num\'eraire (change of currency) technique when the underlying is only a local martingale modelling for example an exchange rate.  The new pricing operator assigns prices to contingent claims according
to the minimal cost for superreplication strategies that succeed with
probability one for both currencies as num\'eraire. 
Within this context, we interpret the lack of the martingale property of an exchange-rate
as a reflection of the possibility that the num\'eraire currency may devalue
completely against the asset currency (hyperinflation).
\end{abstract}
\section{Introduction}
We propose to modify the notion of a contingent claim price in  the setting where the source of uncertainty is a strict local martingale rather than a martingale. More precisely, we propose to use the minimal cost for superreplicating a given contingent claim under two probability measures simultaneously as a pricing operator for contingent claims. In the case of Foreign Exchange markets with X modelling the exchange rate (for example, the price of one Euro in Dollars) the two measures can be thought of as a ``Dollar measure'' and a ``Euro measure'' corresponding to the choice of Dollars or Euros as num\'eraires. The two measures are not equivalent if X is a strict local martingale (that is, a local martingale that is not a martingale).  In this case, the cost for joint superreplication is higher than the expected value under the local martingale measure.  

Our main result is Theorem~\ref{T minimal}, which provides a formula for the minimum joint superreplication cost in a complete market.  This approach restores put-call parity and international put-call equivalence for model prices, and gives the price $X_0$ for the contingent claim that pays $X_T$ at time $T$.  Our pricing formula agrees with the proposals of other authors (Lewis \cite{Lewis},  Madan and Yor \cite{MadanYor_Ito}, Andersen \cite{Andersen}); the novelty is the rigorous justification of this formula as a hedging cost.

The mathematical contribution of this paper is mainly contained in Section~\ref{S changemeasure} and the appendix.  We show how to construct the measure corresponding to a num\'eraire that is allowed to vanish.  Towards this end, we construct the F\"ollmer measure for nonnegative local martingales, extending the corresponding results for strictly positive local martingales. We also develop a stochastic calculus for the suggested change of measure, in which neither measure dominates the other one. 

Section~\ref{S replication} contains the main financial results of the paper.  Our approach uses two num\'eraires simultaneously, which requires us to reintroduce the notions of market completeness and superreplication.   We introduce a model of the market and define trading strategies and contingent claim replication.  After proving our main result on the minimal replicating price, we give numerous corollaries and examples.

In Section~\ref{SS physical},  we consider a physical measure under which both currencies might completely devalue against the other.  In such a situation an equivalent probability measure cannot exist under which the exchange rate follows local martingale dynamics.  Instead each of the risk-neutral measures is only absolutely continuous with respect to the physical measure.  However, as one may use both currencies as hedging instruments, superreplication of contingent claims might still be possible. We provide a set of conditions under which replicating strategies exist and we show  in Proposition~\ref{P minimal P} how the minimal cost for such a strategy is exactly given by the pricing operator of this paper.

This point of view gives us an interpretation of the lack of martingale property of an exchange rate under a risk-neutral probability measure as the positive probability of complete devaluations of currencies (corresponding to explosions of the exchange rate) occurring under some dominating probability measure. We remark that this dominating probability measure does usually not correspond to the F\"ollmer measure, which we shall discuss below, but is equivalent to the sum of the F\"ollmer measure and the original measure.

\subsection*{Related literature}
We now link our financial results to relevant literature:

	Strict local martingales, that is, local martingales that are not martingales,
have recently been introduced in the financial industry to model exchange rates under the  risk-neutral measure. This is
due to the fact that they are able to capture observed features of the market well such as implied volatility surfaces and that they are easily analytically
tractable. An important example is the class of  ``quadratic normal volatility'' models, a family of local martingales, which for example are studied in Andersen \cite{Andersen} and in our companion paper Carr et al.~\cite{CFR_qnv}. 

		There is a vast literature on pricing options in strict local martingale models, often coined ``bubbles.''  For an overview of this literature, we refer the interested reader  to the recent survey by Protter \cite{Protter_2013_bubbles}. 
		 Heston et al.~\cite{HLW} were among the first to point out that put-call parity usually does not hold in strict local martingale models. For a discussion of these models specifically in the context of Foreign Exchange, we refer the reader to
Jarrow and Protter \cite{JP_foreign}.   

Further models in which strict local martingales appear can be found among the class of stochastic volatility models; Sin \cite{Sin} was among the first to point this out. For example, in the log-normal SABR model, if the asset price process is positively correlated with the stochastic volatility process, then it follows strict local martingale dynamics; see Example~6.1 in Henry-Labord\`ere \cite{HenryLabordere}.
	
	Several papers suggest adjustments to the pricing of contingent claims by expectations in strict local martingale models in order to address the lack of put-call parity:
	
 Lewis \cite{Lewis} proposes to add a correction term to the price of a call. However, this approach lacks a clear economic motivation. As his starting point is exactly put-call parity for model prices, it is not clear how other contingent claims should be priced.   
			
			Cox and Hobson \cite{CH} suggest to consider collateral requirements when pricing contingent claims; such collateral requirements correspond to a constraint on the class of admissible trading strategies. This leads to a higher contingent claim price, but usually does not restore put-call parity for model prices.

Madan and Yor \cite{MadanYor_Ito} propose to take
the limit of a sequence of prices obtained from approximating the asset price by true martingales
as the price for a contingent claim. This approach also restores put-call parity for model prices.
However, one might criticize that
the limit of the approximating prices does usually not agree with the classical price in the case that the underlying is a true martingale. For instance, consider an arbitrage-free, complete market with corresponding risk-neutral measure $\Q$, a standard  $\Q$-geometric Brownian motion $X = \{X_t\}_{t \in [0,T]}$ as underlying, and a contingent claim that pays 
$\1_{\{X_T \in \N\}}  / q_{X_T}$  at maturity $T$ where $q_y =   \Q(\max_{t \in [0,T]} X_t \geq y)$ for all $y \in \R$.  This claim should have price zero as $X_T \notin \N$ almost surely.  However, if one approximates $X$ with versions that are stopped at hitting times of integers, as in Madan and Yor's approach, then one obtains a price of one for that claim.

We here suggest to take an economic point of view based on a replicating argument and derive a
pricing operator that restores put-call parity and, therefore, assigns model prices that correspond to observed market prices.
We thus not only justify Lewis' pricing operator by an economic argument but also generalize it to a wider
class of models and contingent claims. 

The approach taken here can be interpreted as a link between classical pricing and pricing under Knightian uncertainty.  Pricing in the classical sense corresponds to the choice of one probability measure under which a contingent claim is superreplicated. This choice implies a strong assumption on the chosen nullsets, that is, by assumption a set
of events is determined to be not relevant for computing a replicating trading strategy. If the modeler considered
another probability measure, other events would be selected, leading to a different replicating price and
strategy. Indeed, one would like that the choice of probability measure should not have a large impact on the price (or, more importantly, on the hedging strategy) or, at least,
should be quantifiable.



We remark that Delbaen and Schachermayer \cite{DS_numeraire} work out the connection of strict local martingales and changes of num\'eraires.  While we understand a change of num\'eraire as a combination of a change of currency and the corresponding change of measure they start by looking at the change of currency only.  Their results imply that, in an arbitrage-free model, a change of currency leads to the existence of arbitrage if the corresponding exchange rate is a strict local martingale under a unique risk-neutral measure. We circumvent the appearance of arbitrage here by associating the change of num\`eraire with the introduction of a new probability measure that is not equivalent to the old one.  It is exactly this lack of equivalence which avoids the arbitrage after the change of currency.
%
%
\section{Change of measure with a nonnegative local martingale}  \label{S changemeasure}
In this section,  given a probability measure $\Q$, we construct and discuss a new probability measure $\widehat{\Q}$ corresponding to a density process that  follows local martingale dynamics only
and is allowed to hit zero.
It is helpful to interpret the notation of this section in a financial context. Towards this end, we interpret $X$ as an exchange rate, for example, the price of one Euro in Dollars. Then $\Q$ represents the risk-neutral measure corresponding to the Dollar-num\'eraire and $\widehat{\Q}$ the  risk-neutral  measure corresponding to the Euro-num\'eraire.

The Mathematical Finance literature
has utilized the techniques developed by F\"ollmer \cite{F1972} and Meyer \cite{M} to  construct probability measures with a strict local martingale as density process; mostly in the context 
of arbitrage and bubbles; see, for example, Delbaen and Schachermayer \cite{DS_Bessel}, Pal and Protter \cite{PP}, Fernholz and Karatzas \cite{FK}, Ruf \cite{Ruf_hedging}, and, parallel to this work, Kardaras et al.~\cite{KKN_local}.
Here, we slightly extend this literature by allowing the local martingale to hit zero.
On the other side, true martingales, possibly hitting zero, as density processes have been studied by Sch\"onbucher \cite{Schoenbucher_2000} within the area of Credit Risk.
Sch\"onbucher \cite{Schoenbucher_2000} terms the corresponding measure a ``survival measure.'' We extend this direction of research by allowing the change of measure  being
determined by a local martingale only.

Throughout this section, we fix a time horizon $T \in (0,\infty]$, a stochastic basis $(\Omega, \CF_T, \{\CF_t\}_{t \in [0, T]}, \Q)$, and a nonnegative $\Q$-local martingale $X = \{X_t\}\tInd$. We assume that $x_0 := X_0>0$ is deterministic, that $\{\CF_t\}_{t \in [0, T]}$ is right-continuous, and that $X(\omega)$ has right-continuous paths for all $\omega \in \Omega$; see also Lemma~1.1 in F\"ollmer \cite{F1972} for the construction of a right-continuous version if this assumption does not hold. 

Any nonnegative random variable, such as $X_T$, is explicitly allowed to take values in $[0,\infty]$.  For a
nonnegative random variable $Z$ and some set $A \in \CF_T$,
we will write $Z \1_A$ to denote the random variable that equals $Z$
whenever $\omega \in A$, and otherwise is zero. 
 For any stopping time $\tau$ we shall denote the stochastic process
that arises from stopping a process $N = \{N_t\}\tInd$ at time
$\tau$ by $N^\tau = \{N_t^\tau\}\tInd;$ that is, $N^\tau_t := N_{t
\wedge \tau}$ for all $t \in [0,T].$  For any measure $\Prob$ on $(\Omega, \CF_T)$, we denote the corresponding expectation operator by $\E^{{\Prob}}$.

We define the stopping times
\begin{align*}
    R_i &:= \inf\{t \in [0,T]: X_t > i\},   \\
    S_i &:= \inf\left\{t \in [0,T]: X_t < \frac{1}{i}\right\}  
\end{align*}
for all $i \in \N$, $R := \lim_{i \uparrow \infty} R_i$, and $S := \lim_{i \uparrow \infty} S_i$,
with the convention that $\inf \emptyset := \mathfrak{T}$ for some trans-finite time $\mathfrak{T} > \infty$; see Appendix~\ref{A stoch interval} for details.
We define the process $Y = \{Y_t\}\tInd$ by $Y_t := 1/X_t \1_{\{ R> t\}}$ for all $t \in [0,T]$ and the stopping times
\begin{align*}
    S_i^Y &:= \inf\left\{t \in [0,T]: Y_t > i\right\}, 
\end{align*}
again with $\inf \emptyset := \mathfrak{T}$. We observe that $S_i = S_i^Y$ for all $i \in \N$ if $X_{(R+t) \wedge T} = \infty$ for all $t \geq 0$.
We shall assume that  $\{\CF_t \cap \CF_{R-} \}_{t \in [0, T]}$ is the right-continuous modification of a standard system; see Appendix~\ref{A conditions} for notation and a discussion of this assumption.

The next theorem states the main result of
this subsection; for the nonnegative $\Q$-local martingale $X$ there exists a probability measure under which
$X$ serves as the num\'eraire. We remark that we only specify the new measure on $(\Omega, \CF_{R-})$ and not on $(\Omega, \CF_T)$. This is due to 
the fact that the original measure $\Q$, by assumption, does not ``see'' any events after the stopping time $R$. However, a measure on
$(\Omega, \CF_T)$ satisfying the properties of the next theorem could be easily constructed by arbitrarily, but
consistently, extending the measure $\widehat{\Q}$ from $\CF_{R-}$ to $\CF_T$. 
Observe that $Z \1_{\{R  > \tau \wedge T\}}$ is $\CF_{R_-}$--measurable if $Z$ is an $\CF_{\tau \wedge T}$--measurable random variable for some stopping time $\tau$.

\begin{theorem}[Change of measure with a nonnegative local martingale]  \label{T numeraire}
    There exists a unique probability measure $\widehat{\Q}$ on $(\Omega, \CF_{R-})$
	such that
    \begin{align} \label{E Qe0}
        \widehat{\Q}\left(A \cap \left\{R  > \tau \wedge T\right\}\right) = \frac{\E^{\Q}\left[\1_A X^\tau_T\right]}{x_0}
    \end{align} 
    holds for all stopping times $\tau$ and $A \in \CF_{\tau \wedge T}$.    This measure also  satisfies $\widehat{\Q}(R_i \wedge T<R) = 1$ for all $i \in \N$. Moreover, we have 
    \begin{align} 
        \E^{\widehat{\Q}}\left[Z \1_{\{R  > \tau \wedge T\}}\right] &= \frac{\E^{\Q}\left[(Z \1_{\{S>\tau \wedge T\}}) X^\tau_T\right]}{x_0} \label{E Qe}
      \end{align}
     and
     \begin{align}
        \E^{{\Q}}\left[Z \1_{\{S  > \tau \wedge T\}}\right] &= x_0 \E^{\widehat{\Q}}\left[(Z \1_{\{R>\tau \wedge T\}}) Y^\tau_T\right] \label{E QdQe}
    \end{align}
        for all stopping times $\tau$ and $\CF_{\tau \wedge T}$--measurable random variables $Z \in [0,\infty]$. 
   
The process $Y$ satisfies $$\widehat{\Q}(\inf_{t \in [0,T]} \{Y_t\} \geq 0) = 1$$; furthermore,  we have $$\widehat{\Q}(\inf_{t \in [0,T]} \{Y_t\} > 0) = 1$$  if and only if the process $X$ is a (uniformly integrable) $\Q$-martingale.  Moreover:
	\begin{enumerate}
		\item $Y$ is a $\widehat{\Q}$-supermartingale;
		\item $Y$ is a local $\widehat{\Q}$-martingale if and only if  ${\Q}(S>S_i \wedge T) = 1$ for all $i \in \N$; then $\{S_i^Y\}_{i \in \N}$ is, under $\widehat{\Q}$, a localization sequence for $Y$;
		\item $Y$ is a $\widehat{\Q}$-martingale if and only if ${\Q}(S > T) = 1$.
	\end{enumerate}
\end{theorem}
\begin{proof}
Without loss of generality, assume  throughout the proof that $x_0 = 1$.
	Observe 
	that $X^{R_i}$ is a nonnegative martingale by Lemma~\ref{L locsequence2} in the appendix; thus, it generates a measure $\Q_i$ on  $(\Omega, {\CF}_{R_i-})$  by $\dd \Q_i := X^{R_i}_T \dd \Q$ for all $i \in \N$. Observe that
the family of probability measures $\{\Q_i\}_{i \in \N}$ is consistent, that is, $\Q_{i+j}|_{{\CF}_{R_i-}} = \Q_i$ for all $i,j \in \N$, and that  ${\CF}_{R-} = \bigvee_{i \in \N} {\CF}_{R_i-}$.
Now, the Extension Theorem~V.4.1 in Parthasarathy \cite{Pa} yields the existence of a probability measure $\widehat{\Q}$ on  $(\Omega, {\CF}_{R-})$ such 
that $\widehat{\Q}|_{{\CF}_{R_i-}} = \Q_i$; see Appendix~\ref{A conditions} for an argument that the necessary assumptions of that theorem hold.

Observe that 
    \begin{align*}
	\widehat{\Q}\left(A \cap \left\{R  > \tau \wedge T\right\}\right) 
            &= \lim_{i \uparrow \infty} \widehat{\Q}\left(A \cap \left\{ R_i > \tau \wedge T\right\}\right)
            = \lim_{i \uparrow \infty} \Q_i \left(A \cap \left\{ R_i > \tau \wedge T\right\}\right)\\
            &= \lim_{i \uparrow \infty} \E^{\Q}\left[ \1_{A \cap \left\{ R_i > \tau \wedge T\right\}} X^{R_i}_T\right]
            = \lim_{i \uparrow \infty} \E^{\Q}\left[\1_{A \cap \left\{ R_i > \tau \wedge T\right\}} X^{\tau}_T \right]\\
            &= \E^{\Q}\left[\1_A X^\tau_T \right]
    \end{align*} 
for all $A \in \CF_{\tau \wedge T}$ and stopping times $\tau$.  This yields \eqref{E Qe0}. Now, with $\tau = R_i$ and $A =\Omega$ we obtain that  $\widehat{\Q}\left(R  > R_i \wedge T\right) = 1$ for all $i \in \N$. This identity implies that $\widehat{\Q}(A) = \widehat{\Q}\left(A \cap \left\{R  > R_i \wedge T\right\}\right)  = \E^{\Q}[X^{R_i}_T \1_A ]$ for all $A \in  \CF_{R_i-}$ and $i \in \N$.  
Since ${\CF}_{R-} = \bigvee_{i \in \N} \CF_{R_i-}$ and $\bigcup_{i \in \N} \CF_{R_i-} $ is a $\pi$-system, this yields uniqueness of $\widehat{\Q}$ on $(\Omega, \CF_{R-})$.
Then \eqref{E Qe} follows from \eqref{E Qe0} by applying the monotone convergence theorem and \eqref{E QdQe} follows from \eqref{E Qe} by using the fact that $\Q(R>T)=1$ and formally applying \eqref{E Qe}  to $(Z \1_{\{R > \tau \wedge T\}}) Y^\tau_T$ instead of $Z$, where $\tau$ and $Z$ are as in the theorem.

Observe that $\{\inf_{t \in [0,T]} \{Y_t\} < 0\} = \bigcup_{i \in \N} \{\inf_{t \in [0,R_i)} \{Y_t\} < 0\}$ $\widehat\Q$-almost surely and conclude that $Y$ is $\widehat{\Q}$-almost surely  nonnegative by dominated convergence and applying \eqref{E Qe0} with $\tau = R_i$ and $A = {\{\inf_{t \in [0,R_i)} \{Y_t\} < 0\} }$ for all $i \in \N$. 
Next,  \eqref{E Qe0}  with $\tau = T$ and $A = {\{\inf_{t \in [0,T]} \{Y_t\} > 0\} }$ yields that $\widehat{\Q}(\{\inf_{t \in [0,T]} \{Y_t\} > 0\} ) = \E^\Q[X_T]$, which shows the equivalence of the positivity of $Y$ under $\widehat{\Q}$ and the martingale property of $X$ under $\Q$.

	Observe that
    \begin{align*}  
        \E^{\widehat{\Q}}\left[Y_t \1_A\right] =
            \E^{\widehat{\Q}}\left[Y_{t} \1_A \1_{\{R > t\}}\right] = 
            \Q\left(A \cap \left\{S >t\right\} \right) 
            \leq  \Q\left(A \cap \left\{S > s\right\} \right)  =\E^{\widehat{\Q}}\left[Y_s \1_A\right] 
    \end{align*}
	by applying \eqref{E QdQe}  twice with $Z = \1_A$ and $\tau =  t$ the first time and $\tau = s$ the second time for all $t \in [0,T]$, $s \in [0,t]$, and $A \in \CF_s$. Thus, the process $Y$ is a $\widehat{\Q}$-supermartingale.
	This implies, for all stopping times $\tau$, that $Y^\tau$ is a $\widehat{\Q}$-martingale if and only if $\Q\left(S > \tau \wedge T\right) = 1$  since  $\Q\left(S > \tau \wedge T\right)  = \E^{\widehat{\Q}}\left[Y_T^\tau\right]$ again by \eqref{E QdQe} with $Z = 1$.  Using $\tau = T$ yields that $Y$ is a $\widehat{\Q}$-martingale if and only if $\Q\left(S >T\right) = 1$ and using $\tau = S^Y_i$ yields that, for all $i \in \N$,   $Y^{S_i^Y} $ is a $\widehat{\Q}$-martingale if and only if $\Q\left(S >S_i \wedge T\right)  = 1$ since $\Q(S_i^Y = S_i)=1$. We conclude by applying (i) and the equivalence of (a.1) and (a.2) in   Lemma~\ref{L locsequence2} in the appendix.
\qed\end{proof}

    It is important to note that the two measures $\Q$ and $\widehat{\Q}$ are usually not absolutely continuous with
    respect to each other; in particular, it is possible
    that $\widehat{\Q}(R \leq T) > 0 = \Q(R \leq T)$.  
    Furthermore, note that the indicators in \eqref{E Qe} and \eqref{E QdQe} can be omitted if
    $Z$ is finite or $\Q(S \leq T) = 0$ or $\widehat{\Q}(R \leq T) = 0$, respectively.
    In general, however, the indicators are necessary as the example $Z=1/X^S_T$ illustrates.
\begin{remark}[Duality of martingale property and  positivity of density processes] \label{R duality}
	Observe that we proved the equivalence of the following statements in Theorem~\ref{T numeraire}:
\begin{enumerate}
	\item $Y$ is a (uniformly integrable) $\widehat{\Q}$-martingale on $[0,T]$;
	\item $\Q(S>T) = 1$.
\end{enumerate}
	We also proved the equivalence of the following statements:
\begin{enumerate}
	\item $X$ is a (uniformly integrable) ${\Q}$-martingale on $[0,T]$;
	\item $\widehat{\Q}(R>T) = 1$.   
\end{enumerate}
We emphasize the symmetry of these two equivalences.

This duality of the martingale property of a nonnegative local martingale under one measure and its non-explosiveness under another measure has been utilized to provide conditions for the martingale property of local martingales; see, in particular,  Cheridito et al.~\cite{CFY} and Ruf \cite{Ruf_martingale, Ruf_Novikov}.
 \qed
\end{remark}

We next derive properties of the change of measure  in Theorem~\ref{T numeraire}. In
particular, we focus on understanding which of the martingale properties of stochastic processes survive
the change of measure, possibly after modifying the processes. 
The discussion here involves local martingales on stochastic intervals. This notion generalizes the definition of local martingales; its precise definition can be found in Appendix~\ref{A stoch interval}. 

\begin{proposition}[Equivalence of (local) martingales]  \label{P Equivalence}
	Assume the notation of Theorem~\ref{T numeraire}  and let  $\tau$ denote a stopping time and  $N = \{N_t\}\tInd$ a  progressively measurable stochastic process taking values in $[0,\infty]$ such that $N_t = N_t \1_{\{R>t\}}$ for all $t \in [0,T]$.
    The following statements then hold:
    \begin{enumerate}
        \item[(i)] The process $\{N^\tau_t \1_{\{S > \tau \wedge t\}}\}\tInd$ is a $\Q$-martingale if and only if ${N}^\tau Y^\tau$ is a $\widehat{\Q}$-martingale.
        \item[(ii)] The process  $\{N_t \1_{\{S>t\}}\}\tInd$ is a $\Q$-local martingale on $[0,S)$ (equivalently, on $[0, R \wedge S)$) if and only if ${N Y}$ is a $\widehat{\Q}$-local martingale on $[0,R)$  (equivalently, on $[0, R \wedge S)$).
        \item[(iii)] If $\{N_t^{S_i} \1_{\{S>S_i \wedge t\}}\}\tInd$ is a $\Q$-martingale for all $i \in \N$ then  ${N} Y$ is a $\widehat{\Q}$-local martingale. 																																			
        \end{enumerate}
\end{proposition}
The proof of Proposition~\ref{P Equivalence} is based on an extended version of Bayes' formula and can be found in Appendix~\ref{A proofs}.
Applying $(i)$ to $N \equiv 1$ and to $N = X$ with $\tau = T$ yields exactly the equivalences of Remark~\ref{R duality}. Applying (ii) to $N=1$ yields that $Y$ is a $\widehat{\Q}$-local martingale on $[0,R \wedge S)$ if and only if $S$ is announced under $\Q$ (for example, if $X$ does not jump to zero under $\Q$).  The example $X = Y\equiv 1$ shows that the reverse direction in (iii) in Proposition~\ref{P Equivalence} usually does not hold.
%

In order to better understand the suggested change of measure in this section, it is instructive to study an
extreme case where the measures $\Q$ and $\widehat{\Q}$ are not only not absolutely continuous with respect to each other
but even singular:
\begin{example}[Singular measures]  \label{example singularI}
	Assume that $T \in (0,\infty)$ and let $X$ be defined by
    \begin{align} \label{E singular X}
        X_t = 1+\int_0^{t\wedge \widetilde{S}} \frac{1}{\sqrt{T-u}} \dd W_u
    \end{align}
    for all $t \in [0,T)$, where $W = \{W_t\}\tInd$ denotes a $\Q$-Brownian motion and $\widetilde{S} = S$ the first hitting time of $-1$ by $\{\int_0^t 1/\sqrt{T-u} \dd W_u\}_{t \in [0,T)}$.
    Since $X$ corresponds to a
    deterministically time-changed
    Brownian motion, we have  $\Q(S < T) = 1$ and thus $\Q(X_T = 0) =1$.
    
    Under the measure $\widehat{\Q}$ of Theorem~\ref{T numeraire}, note
    that $Y = \{Y_t\}\tInd$, defined by $Y_t := 1/X_t \1_{\{R>t\}}$ for all $t \in [0,T]$, has the dynamics
    \begin{align} \label{E singular Y}
        \dd Y_t = -Y_t^2 \frac{1}{\sqrt{T-t}} \dd W_t^{\widehat{\Q}}
    \end{align}
    for all $t \in [0,R)$ and some $\widehat{\Q}$-Brownian motion
    $W^{\widehat{\Q}} := \{W_t^{\widehat{\Q}}\}_{(t \in [0,R))}$. 
Thus,  $Y$ is just the
    time-change of the reciprocal of a three-dimensional $\widehat{\Q}$-Bessel process $Z$ starting in one. To see this, define the processes
    $Z = \{Z_u\}_{u \geq 0}$ and $B = \{B_u\}_{u \geq 0}$ by $Z_u := Y_{T(1-\exp(-u))}$ and
    \begin{align*}
        B_u := \int_0^{T(1-\exp(-u))} \frac{1}{\sqrt{T-v}} \dd W_v^{\widehat{\Q}}
    \end{align*}
    for all $u \geq 0$. Then observe that $\dd Z_u = -Z_u^2 \dd B_u$ and $\langle B\rangle_u = u$ for all $u \geq 0$ and conclude by applying  L\'evy's theorem; see Theorem~3.3.16 in Karatzas and Shreve \cite{KS1}. We then obtain that
    $Y_t = Z_{\log(T/(T-t))}$ and $\widehat{\Q}(Y_t > 0 \text{ for all }t \in [0,T)) = 1 = \widehat{\Q}(Y_T = 0) $; see
    Section~3.3 of  Karatzas and Shreve \cite{KS1}. Indeed, note that $\widehat{\Q}(R=T)  = \lim_{t \uparrow T; t <T} \widehat{\Q}(R>t) - \widehat{\Q}(R>T)= 1$.
    
    Thus, the two measures are singular with respect to each other on $\CF_T$ since
    $\Q(R=T) = 0 < 1 = \widehat{\Q}(R=T)$;
    however, $\widehat{\Q}$ is absolutely continuous  with respect to $\Q$ on $\CF_t$
    for all $t \in [0,T)$ since $Y$ is a strictly positive, strict $\widehat{\Q}$-local martingale;
    see Remark~\ref{R duality}. We also note that $X^t$ is a true $\Q$-martingale for all $t \in [0,T)$,
    but $X=X^T$ is a strict $\Q$-local martingale.
    \qed
\end{example}
The next example is a slight modification of the example in Delbaen and Schachermayer
\cite{DS_counter}. It  here illustrates that the equivalence of two
probability measures $\Q$ and $\Q^Z$ on $(\Omega,
\CF_T)$, under which $X$ is a nonnegative right-continuous local martingale, does not necessarily imply the equivalence of the
corresponding probability measures $\widehat{\Q}$  and
$\widehat{\Q}^{Z}$, constructed as in Theorem~\ref{T numeraire}. This observation will be one reason why we
shall assume complete markets later on.
\begin{example}[Lack of equivalence]  \label{example LackEquiv}
Fix $T \in (0,\infty)$ and let $\mathcal{X} = \{\mathcal{X}_t\}\tInd$ and $\mathcal{Z} = \{\mathcal{Z}_t\}\tInd$ denote two independent processes with the same distribution as the process $X$
 in Example~\ref{example singularI}. Define
    the stopping time
    \begin{align*}
        \tau^{\mathcal{X} } := \inf\left\{t \in [0,T]: \mathcal{X}_t < \frac{1}{2}\right\}
    \end{align*}
    with $\inf \emptyset := \mathfrak{T}$, and similarly, $\tau^{\mathcal{Z} }$.  Define now the processes $X := \mathcal{X}^{\tau^{\mathcal{X} }  \wedge \tau^{\mathcal{Z} } }$ and $Z := \mathcal{Z}^{\tau^{\mathcal{X} }  \wedge \tau^{\mathcal{Z}}}$. Since the stopping time $\tau^{\mathcal{Z} }$ is independent from $\mathcal{X}$ and satisfies $\Q(\tau^{\mathcal{Z} } < T) = 1$, the process $X$ is a strictly positive true $\Q$-martingale by a conditioning argument; and similarly, so is $Z$.

    Define now a new probability measure $\Q^Z$ by $\dd\Q^Z = Z_T \dd \Q$ and observe that $\Q$ and $\Q^Z$ are equivalent and that
    the process  $X$ is a strict $\Q^{Z}$-local martingale since
    \begin{align*}
        \E^{\Q^{Z}}[X_T] &= \E^{\Q}[X_T Z_T] = \frac{1}{2} \E^{\Q}\left[X_T \1_{\{\tau^{\mathcal{Z} }< \tau^{\mathcal{X} }\}} + Z_T \1_{\{\tau^{\mathcal{X} }< \tau^{\mathcal{Z} }\}}\right]\\
	& =\E^{\Q}\left[X_T \1_{\{\tau^{\mathcal{Z} }< \tau^{\mathcal{X} }\}} \right] =1-\E^{\Q}\left[X_T \1_{\{\tau^{\mathcal{X} }< \tau^{\mathcal{Z} }\}} \right] = 
	1 - \frac{1}{2}\Q(\tau^{\mathcal{X} }< \tau^{\mathcal{Z} }) \\&= \frac{3}{4} < 1, 
    \end{align*}
	where we used the definitions of $\Q^Z$, $ \tau^{\mathcal{X} }$, and  $\tau^{\mathcal{Z} }$, the fact that $\Q(  \tau^{\mathcal{X} }=\tau^{\mathcal{Z} }) =0$ and that $X$ and $Z$ have the same distribution under $\Q$, and the martingale property of $X$ under $\Q$.

    Let $\widehat{\Q}$  and $\widehat{\Q}^Z$
    now denote the probability measures of Theorem~\ref{T
    numeraire} with $X$ as density process. These two measures
    cannot be equivalent since $X$ is a strictly positive true
    $\Q$-martingale, but only a strict $\Q^Z$-local martingale. Thus, the measure $\Q$, $\Q^Z$, and $\widehat\Q$ are all equivalent but only absolutely continuous with respect to $\widehat\Q^Z$.

    To elaborate on this, under both measures $\widehat{\Q}$  and
    $\widehat{\Q}^Z$, the process $1/X$ is a martingale and follows the same dynamics
    as the process $Y$ in \eqref{E singular Y}, stopped at time
    $\tau := {\tau}^{\mathcal{X} } \wedge {\tau}^{\mathcal{Z} } $. However,
    the distribution of ${\tau} $ varies under the two
    measures. More precisely,
    under $\widehat{\Q}$, the stopping time $\tau \leq {\tau}^{\mathcal{Z} }$ is bounded by the first time
    that the nonnegative $\widehat{\Q}$-local martingale $\mathcal{Z}$ starting in $1$ hits
    $1/2$; thus $\widehat{\Q}({\tau} <T) =1$; see also (ii) of Proposition~\ref{P Girsanov} in the appendix.
However, both $1/X$ and $1/Z$ are $\widehat{\Q}^Z$-martingales and the event that neither of these two  $\widehat{\Q}^Z$-martingales hits $2$ has positive probability under $\widehat{\Q}^Z$; thus $\widehat{\Q}^Z(\tau > T) > 0$.
    This yields that $\widehat{\Q}(R \leq T) = 0 <
    \widehat{\Q}^Z(R \leq T)$, despite ${\Q}$  and
    ${\Q}^Z$ being equivalent.
    \qed
\end{example}
\section{Minimal joint replication price}   \label{S replication}
In this section,  we derive and discuss a representation of a contingent claim price, which we define as the minimal replicating cost of the contingent claim's payoff under two probability measures simulaneously; specifically the measure under which the underlying follows local martingale dynamics ($\Qd$) and the measure that corresponds to  the change of num\'eraire  ($\Qe$).  We interpret a nonnegative $\Qd$-local martingale $X$ as the current market value of one Euro in Dollars, and the process $Y:=1/X$ under the measure $\Qe$, derived from $\Qd$ via the density process $X$ (see Theorem~\ref{T numeraire}), as the current market value of one Dollar in Euros under its corresponding num\`eraire measure.

\begin{remark}[Arbitrage and strict local martingales]
Modelling asset prices with strict local martingales usually leads to features of contingent claim prices  that, on the first look, seem to imply simple  arbitrage opportunities and do not reflect our economic
understanding of financial markets. To elaborate on this issue more, we remind the reader of the
standard definition of a contingent claim price in a complete market framework as the minimal (super-) replicating cost of this contingent claim; here the replication occurs almost surely under the unique risk-neutral measure.

Using such contingent claim prices then usually results in the loss of standard put-call parity  in strict local martingale models; see, for example, Cox and Hobson \cite{CH}.
Even more disturbingly, the minimal replicating price for an asset modelled as a strict local martingale in a
complete market is below its current value. Yet, due to an admissibility
constraint on trading strategies, these models do not yield arbitrage opportunities; see also Delbaen and Schachermayer \cite{DS_fundamental, DS_1998}. For example, the strategy of shorting the asset modelled by a strict local martingale and replicating its payoff for a lower cost is not admissible, as it might lead to unbounded negative wealth before the strategy matures; more details on this argument are discussed in Ruf
\cite{Ruf_negative}.
\qed
\end{remark}

Throughout this section, we again assume a time horizon $T \in (0,\infty]$ and a filtered probability space $(\Omega, \CF_T, \{\CF_t\}\tInd, \Qd)$ that satisfies the  technical conditions of Appendix~\ref{A conditions}.
We fix a nonnegative $\Qd$-local martingale $X$ with almost surely \cadlag{} paths (and right-continuous for all $\omega \in \Omega$), define the stopping times $\{R_i\}_{i \in \N}$, $\{S_i\}_{i \in \N}$, $S$, and $R$ as in  Section~\ref{S changemeasure}, and assume that $\Qd(S>S_i \wedge T) = 1$ for all $i \in \N$; that is, $X$ is assumed not to jump to zero. As above, we define a process $Y = \{Y_t\}\tInd$ by $Y_t := 1/X_t \1_{\{ R > t\}}$ for all $t \in [0,T]$.
As illustrated in Theorem~\ref{T numeraire}, there exists a probability measure $\Qe$, which corresponds
to the probability measure with $X$ as num\'eraire, symbolically $\text{``} \dd \Qe = X_T \dd \Qd\text{''}$.   We then extend the measure  $\Qe$, currently defined on $(\Omega, \CF_{R-})$, to a measure on $(\Omega, \CF_T)$, which we again denote, with a slightly misuse of notation, by $\Qe$; see Appendix~\ref{A conditions}. 

For some $d \in \N$, 
we assume the existence of $d+1$ tradable assets with nonnegative \cadlag{} price processes (right-continuous for all $\omega \in \Omega$), denoted by $S^{\$} = \{S^{\$, (i)}\}_{i = 0, 1, \ldots d}$ in Dollars and by $S^{\eu} =\{S^{\eu, (i)}\}_{i = 0, 1, \ldots d}$ in Euros, respectively, with
$S^{\$, (i)} = \{S^{\$, (i)}_t\}\tInd$ and $S^{\eu, (i)} = \{S^{\eu, (i)}_t\}\tInd$ for all $i = 0, 1, \ldots, d$. We assume that the processes $S^{\$, (i)}$ have $\Qd$-local martingale dynamics and $S^{\eu, (i)}$ have $\Qe$-local martingale dynamics for all $i = 0, 1, \ldots, d$.   Moreover, we assume that these price processes denote the same assets and are consistent; that is, we assume that $S^{\eu, (i)}_t \1_{\{R \wedge S>t\}} = S^{\$, (i)}_t Y_t \1_{\{R \wedge S>t\}}$ for all $t \in [0,T]$ and $i = 0, 1, \ldots, d$. Thus, given a Dollar price process $S^{\$, (i)}$ for the $i^\text{th}$ asset, the process $S^{\eu, (i)}$ denotes the price for the same asset in Euros for all $i = 0,1, \ldots, d$, as it is the dollar price multiplied by the price of one Dollar in Euros. This relationship holds up to time $R \wedge S$; any event ``beyond'' that stopping has probability zero under one of the two measures.

By Proposition~\ref{P Equivalence} and by Proposition~\ref{P extension} in the appendix, given the $\Qd$-dynamics of $S^{\$}$, we can always construct $\Qe$-local martingales $S^{\eu}$ with $S^{\$, (i)}_t \1_{\{R \wedge S>t\}} = S^{\$, (i)}_t Y_t \1_{\{R \wedge S>t\}}$ for all $t \in [0,T]$ and $i = 0, 1, \ldots, d$. More precisely, the equivalence in (ii) of Proposition~\ref{P Equivalence} and our standing assumption that $\Qd(S>S_i \wedge T)=1$ for all $i \in \N$ first yield that $S^{\$, (i)} Y$ is a $\Qe$-local martingale on $[0,R \wedge S)$. Secondly, since $R \wedge S$ is foretellable  under  $\Qe$, by Theorem~\ref{T numeraire},  any $\Qe$-local martingale on $[0, R \wedge S)$ can be extended to a local martingale on $[0, R \wedge S]$,  and then, of course, to a local martingale on $[0,T]$ in an arbitrary manner after that time since the dynamics under one measure only determine the dynamics under the other measure up to the stopping time $R \wedge S$.

We suppose that $S^{\$, (0)}$ and $S^{\eu, (1)}$ denote the Dollar and Euro money market account, each assumed to pay zero interest; that is, $S^{\$, (0)} \equiv 1 \equiv S^{\eu, (1)}$. Thus, $S^{\$, (1)} = X$ denotes the price of one Euro in Dollars and $S^{\eu, (0)} = Y$ the price of one Dollar in Euros.  More generally, part  (iii) of Proposition~\ref{P Equivalence} yields that if $S^{\$, (i)}$ is a $\Qd$-martingale for some $i = 1, \ldots, d$ then $S^{\eu, (i)}_t \1_{\{R \leq t\}} = 0$ for all $t \in [0,T]$; to wit,  the martingale property of $S^{\$, (i)}$ under $\Qd$ forces $S^{\eu, (i)}$ to hit zero under $\Qe$ at time $R \wedge S$. Vice versa, if $S^{\eu, (i)}$ is a $\Qe$-martingale for some $i = 1, \ldots, d$ then $S^{\$, (i)}_t \1_{\{S \leq t\}} = 0$ for all $t \in [0,T]$.

We now are ready to define a trading strategy, relying on stochastic integrals with respect to the $d+1$-dimensional local martingales $S^{\$}$ and $S^{\eu}$.  We refer to Sections~I.4d and III.4a in Jacod and Shiryaev \cite{JacodS} for a discussion of stochastic integrals, when the filtration does not satisfy the ``usual assumptions,'' in the case of $d =1$ or all price processes being continuous and to Jacod \cite{Jacod_1980}
for the general case. We denote by $L(S^\$)$ and $L(S^\eu)$ the space of all predictable processes that are integrable with respect to $S^\$$ and $S^\eu$, respectively, under the corresponding measures $\Qd$ and $\Qe$.

   Stochastic integration is used in the following definition:
\begin{definition}[Trading strategy]  \label{D trading}
    A \emph{trading strategy}
    is an $\R^{d+1}$-valued  process $\eta \in L(S^\$) \cap L(S^\eu)$  such that
    \begin{itemize}
        \item its corresponding Dollar wealth process $V^{{\$}, \eta} = \{V^{{\$}, \eta}_t\}\tInd$
            and Euro wealth process $V^{{\eu}, \eta} = \{V^{{\eu}, \eta}_t\}\tInd$,
            defined by
                \begin{align*}
                    V^{{\$}, \eta}_t := \sum_{i=0}^d \eta^{(i)}_t S^{{\$},(i)}_t \quad \text{ and } \quad
                    V^{{\eu},\eta}_t := \sum_{i=0}^d \eta^{(i)}_t S^{{\eu},(i)}_t 
                \end{align*}
            for all $t \in [0,T]$, stay nonnegative almost surely under the corresponding measure
            $\Qd$ and $\Qe$, respectively, and
        \item the self-financing condition holds, that is,
    \begin{align*}
                \dd V^{{\$}, \eta}_t = \sum_{i=0}^d \eta^{(i)}_t \dd S^{{\$},(i)}_t, \quad \text{ and } \quad
                \dd V^{{\eu},\eta}_t = \sum_{i=0}^d \eta^{(i)}_t \dd S^{{\eu},(i)}_t
    \end{align*}
    for all $t \in [0,T]$, where the dynamics are computed under the
    corresponding measure $\Qd$ and $\Qe$, respectively.
        \end{itemize}
    We shall say that $\eta$ is a trading strategy for initial capital $v \in [0,\infty)$
    expressed in Dollars if
    \begin{align*}
        v = V^{{\$}, \eta}_0 = \sum_{i=0}^d \eta^{(i)}_0 S^{{\$},(i)}_0
    \end{align*}
    holds; and similarly for initial capital $v$ expressed in Euros. \qed
\end{definition}
Thus, at any time $t \in [0,T]$, each component of $\eta_t$ determines the current number of
shares of each asset held at that point of time.  Note that
\begin{align}\label{E VV}
	V^{{\eu}, \eta}_t \1_{\{R \wedge S>t\}} = V^{{\$}, \eta}_t  Y_t  \1_{\{R \wedge S>t\}}
\end{align}
 for all $t \in [0,T]$. Thus, the 
 nonnegativity condition on  $V^{{\$}, \eta}$ implies the one on  $V^{{\eu}, \eta}$, but
only up to the stopping time $R$.  Moreover, a simple application of It\^o's rule yields that
the self-financing condition under $\Qd$ implies the one under $\Qe$, but again only up to the stopping time $R$; see also Geman et al.~\cite{Ger}.

We call any pair of nonnegative $\CF_T$--measurable random variables $(D^{\$}, D^{\eu})$ a \emph{contingent claim} if $D^{\eu} \1_{\{R \wedge S>T\}}= D^{\$} Y_T \1_{\{R \wedge S>T\}}$.
The random variable $D^{\$}$ ($D^{\eu}$) corresponds to the Dollar (Euro) price of a contingent claim,
as seen by the Dollar (Euro) investor.  We remind the reader that
the event $\{S \leq T\}$ has zero $\Qe$-probability, but might have
positive $\Qd$-probability, and the converse statement holds for the event $\{R \leq T\}$.

We represent a contingent claim as a pair of random variables in order to be able to exactly express
its payoff both in Dollars and in Euros including in the event of $X$ hitting infinity.
For example, the contingent claim $(X_T,1)$ pays off one Euro at maturity, the contingent claim $(X_T, \1_{\{R > T\}})$
pays off one Euro if the price of one Euro in Dollars did not explode.
For some $K \in \R$,
the claims $D^{C,\$}_K := ((X_T-K)^+, (1-K Y_T)^+)$ and $D^{P,\$}_K := ((K-X_T)^+,(K Y_T-1)^+)$
are called \emph{call} and \emph{put}, respectively, on one Euro with
strike $K$ and maturity $T$. Equivalently, by exchanging the first with the second component and $X_T$ with $Y_T$,
we define calls and puts on one Dollar and denote
them by $D^{C,\eu}_K$ and $D^{P,\eu}_K$.
In Foreign Exchange markets, \emph{self-quantoed calls} are
traded, defined as $D^{SQC,\$}_K := X_T D^{C,\$}_K = (X_T(X_T-K)^+, (X_T-K)^+)$ for some $K \in \R$.

We shall assume that the market is \emph{complete} both for the Dollar investor and for the Euro investor; that is,
for any contingent claim $(D^{\$}, D^{\eu})$ with $D^{\$}, D^{\eu} \in [0,\infty)$
there exist trading strategies $\eta^{\$}$  and $\eta^{\eu}$  such that
\begin{align*}
    \Qd\left(V^{\$, \eta^{\$}}_T = D^{\$}\right) = 1 = \Qe\left(V^{\eu, \eta^{\eu}}_T = D^{\eu}\right)
\end{align*}
and such that $V^{\$, \eta^{\$}}$ is a $\Qd$-martingale and $V^{\eu, \eta^{\eu}}$ is a $\Qe$-martingale.
The replicability of any contingent claim under $\Qd$ does not necessarily imply that any
contingent claim can be
replicated under $\Qe$
since, in general, the two measures are not equivalent. 
%
%


\begin{remark}[A seeming paradox]
    Let the exchange rate $X$ be a strict $\Qd$-local martingale hitting zero with positive
    probability.  Then, $Y$ is a strict $\Qe$-local martingale and
    we have the following paradox. Under the Dollar measure, one can replicate
    the payoff of one Euro for less than one Euro; simultaneously,
    under the Euro measure, one can replicate
    the payoff of one Dollar for less than one Dollar. To conclude, the exchange rate
    reflects an overly-high price (compared to their replicating cost) both for the
    Dollar and for the Euro; thus being at the same time too high and too low for the Dollar.
    This paradox can be explained by reminding oneself that the two measures
    $\Qd$ and $\Qe$ are not equivalent; and therefore the investors are concerned with
    different events when replicating a Euro or a Dollar, respectively. 
    \qed
\end{remark}

The next theorem constitutes the core result of this section; we recall our standing assumption that $\Qd(S>S_i \wedge T) = 1$ for all $i \in \N$:
\begin{theorem}[Minimal joint replicating price]  \label{T minimal}
Define the Dollar and Euro pricing operators as 
    \begin{align}
        p^{\$}(D) &= \E^{\Qd}\left[D^{\$}\right] + x_0 \E^{\Qe}\left[D^{\eu} \1_{\{R \leq T\}}\right], \label{E p}\\
        p^{\eu}(D) &= \E^{\Qe}\left[D^{\eu}\right] + \frac{1}{x_0} \E^{\Qd}\left[D^{\$} \1_{\{S \leq T\}}\right]   = \frac{p^{\$}(D)}{x_0} \label{E p2}
    \end{align}
for a contingent claim $D=(D^\$,D^\eu)$.     
Whenever $D$ is non-negative, the minimal joint $Q^\$$- and $Q^\eu$-replicating price expressed in Dollars (Euros) is $p^\$$ ($p^\eu$).  
More precisely,  there exists some trading strategy $\eta$ for initial capital $p^{\$}(D)$ (expressed in Dollars) such that
    \begin{align}  \label{E minimal}
        \Qd\left(V^{{\$},\eta}_T = D^{\$}\right) = 1 = \Qe\left(V^{{\eu},\eta}_T = D^{\eu}\right);
    \end{align}
    and there exists no $\widetilde{p} < p^{\$}(D)$ and no trading strategy $\widetilde{\eta}$ for initial capital $\widetilde{p}$  (expressed in Dollars) such that
    \eqref{E minimal} holds with $\eta$ replaced by $\widetilde{\eta}$.
\end{theorem}
\begin{proof}
	The second equality in \eqref{E p2} follows directly from Theorem~\ref{T numeraire}.
    Since the market is assumed to be complete there exist trading strategies $\nu$ for initial capital
    $p^{(1)} := \E^{\Qd}[D^{\$}]$ (expressed in Dollars) and $\theta$ for initial capital 
    $p^{(2)} := \E^{\Qe}[D^{\eu} \1_{\{R \leq T\}}]$  (expressed in Euros) such that $V^{{\$},\nu}$ is a $\Qd$-martingale,
    $V^{{\eu},\theta}$ a $\Qe$-martingale, and
    \begin{align*}
        \Qd\left(V^{{\$},\nu}_T = D^{\$}\right) = 1 =
        \Qe\left(V^{{\eu},\theta}_T = D^{\eu} \1_{\{R \leq T\}}\right).
    \end{align*}
    In order to show \eqref{E minimal}, we now prove that the trading stategy $\eta := \nu
    + \theta$ replicates $D^{\$}$ under $\Qd$ and $D^{\eu}$ under $\Qe$;  the initial cost for the strategy $\eta$ is, expressed in Dollars, exactly $p^\$ = p^{(1)} + x_0 p^{(2)}$. Moreover, note the identities
    $V^{{\$},\eta}_T = V^{{\$},\nu}_T + V^{{\$},\theta}_T$ and $V^{{\eu},\eta}_T = V^{{\eu},\nu}_T + V^{{\eu},\theta}_T$.  Therefore, in order to prove that $\eta$ is a trading strategy, it is sufficient to prove that (a) $\Qd(V^{{\$},\theta}_T > 0) = 0$ and (b) $\Qe(V^{{\eu},\nu}_T \1_{\{R \leq T\}} > 0) = 0$. 
    
    For (a), note that $\{V^{\$,\theta}_t  \1_{\{S > t\}}\}_{t \geq 0}$ is a  $\Qd$-local martingale by (i) in Proposition~\ref{P Equivalence} with $N  = \{V^{\$,\theta}_t  \1_{\{R > t\}}\}_{t \geq 0}$ and $\tau = R_i$ for all $i \in \N$. By taking differences, so is $\{V^{\$,\theta}_t  \1_{\{S \leq t\}}\}_{t \geq 0}$, which implies that  $\Qd(\{V^{{\$},\theta}_T > 0\} \cap \{S \leq T\}) = 0$. Observe next that $$\Qd(\{V^{{\$},\theta}_T > 0\} \cap \{S \wedge R_i> T\} ) = x_0 \E^{\Qe}[\1_{\{V^{{\$},\theta}_T > 0\} \cap \{R_i> T\}} Y_T ] = 0$$ by \eqref{E QdQe} and \eqref{E VV} for all $i \in \N$, which yields (a).
    
	For (b), it is sufficient to show that $\{V^{{\eu},\nu}_t \1_{\{R \leq t\}}\}\tInd$ is a (nonnegative) $\Qe$-local martingale.  Since $V^{{\eu},\nu}$ is one, it only remains to show, by \eqref{E VV}, that $NY$ with
	$N := \{V^{{\$},\nu}_t \1_{\{R > t\}}\}\tInd$ is also a $\Qe$-local martingale. However, this follows directly from (iii) in Proposition~\ref{P Equivalence} since $N$ was assumed to be a $\Qd$-martingale.

    Next, we show that $\eta$ corresponds to the cheapest trading strategy. Towards this end, let $\widetilde{p} \in [0,\infty)$ and $\widetilde{\eta}$ be a trading strategy for initial capital $\widetilde{p}$ (expressed in Dollars) that
    superreplicates $D^\$$ under $\Qd$ and $D^\eu$ under $\Qe$.  Then, $\widetilde{p} = M_0 + N_0$, where $M$ and $N$ are the
    martingale and strict local martingale part of the Riesz decomposition $V^{{\$},\widetilde{\eta}} = M + N$ under $\Qd$
    with $\Qd(N_T = 0) = 1$; to wit,
    $M_t = \E^{\Qd}[V^{{\$},\widetilde{\eta}}_T | \CF^0_t]$ and $N_t := V^{{\$},\widetilde{\eta}}_t - M_t$ for all $t \in [0,T]$; see
    Theorem~2.3 of F\"ollmer \cite{F1973} for the case of a not completed filtration.  
    
    Note that $M = V^{{\$},\widetilde{\nu}}$ and  $N = V^{{\$},\widetilde{\theta}}$ for some trading strategies  $\widetilde{\nu}$ and $\widetilde{\theta}$ with $\widetilde{\eta} = \widetilde{\nu} + \widetilde{\theta}$.
Since $\widetilde{\nu}$ superreplicates
    $D^{\$}$ under $\Qd$ we obtain $M_0 \geq \E^{\Qd}[D^{\$}]$. As in (b) in the first part of the proof, we have
    $\Qe(\{M_T > 0\} \bigcap \{R \leq T\}) = 0$. Thus, $\widetilde{\theta}$ superreplicates $D^{\eu}$ under $\Qe$.
   This implies that $N_0 \geq x_0 \E^{\Qe}[D^{\eu} \1_{\{R \leq T\}}]$, which yields that
    $\widetilde{p} = M_0 + N_0 \geq p^{\$}(D)$.
\qed\end{proof}

The last theorem yields the smallest amount of Dollars (Euros) needed to superreplicate a claim $D$ under both measures $\Qd$ and $\Qe$. The corresponding replicating strategy is, as the proof illustrates, a sum of two components. The first component is the standard strategy that replicates the claim under one of the two measures; the second component replicates the claim under the events that only the other measure can ``see.'' 

The next few corollaries are direct implications of the last theorem. We usually formulate
them only in terms of the Dollar pricing operator $p^{\$}$ but symmetrically they also
hold for the Euro pricing operator $p^{\eu}$.
\begin{corollary}[Linearity of pricing operator]
The pricing operator $p^\$$ of \eqref{E p} is linear on its domain: for any claims $D_1 = (D_1^\$,D_1^\eu)$ and $D_2=(D_2^\$,D_2^\eu)$ and any $a \in \R$ such that $D_1$ and  $D_2$ are both in the domain of $p^\$$, we have
    \begin{align*}
        p^{\$}(D_1 + a D_2) = p^{\$}(D_1) +  a p^{\$}(D_2),
    \end{align*}
    where $D_1 + a D_2 := (D_1^{\$} + a D_2^{\$}, D_1^{\eu} + a D_2^{\eu})$.
\end{corollary}
\begin{proof}
    The statement follows directly from the linearity of expectations.
\qed\end{proof}
\begin{corollary}[Martingale property of wealth process]
    The wealth process $V^{{\$},\eta}$ of Theorem~\ref{T minimal} is
    a $\Qd$-local martingale and, thus, does not introduce an arbitrage opportunity. It is
    a strict $\Qd$-local martingale if and only if the $\Qd$-local martingale 
    $X$ is a strict $\Qd$-local martingale and $$\Qe\left(\{D^{\eu} > 0\} \bigcap {\{R \leq T\}}\right) > 0.$$
    Similarly, the wealth process $V^{{\eu},\eta}$ is a $\Qe$-local martingale.
    It is a strict $\Qe$-local martingale if and only if the  $\Qe$-local martingale 
    $Y$ is a strict $\Qe$-local martingale and $$\Qd\left(\{D^{\$} > 0\} \bigcap {\{S \leq T\}}\right) > 0.$$
\end{corollary}
\begin{proof}
    The local martingale property of the wealth processes under the corresponding measures
    follows directly from their definition.
    The lack of martingale property follows from checking when $p^{\$}(D)$ and $p^{\eu}(D)$  in \eqref{E p} and \eqref{E p2} satisfy $p^{\$}(D) > E^{\Qd}[D^{\$}]$ and
    $p^{\eu}(D) > E^{\Qe}[D^{\eu}]$, respectively.
\qed\end{proof}
\begin{corollary}[Price of a Euro]
    The minimal joint $\Qd$- and $\Qe$-super\-replicating price of $(X_T,1)$ is $x_0$ (expressed
    in Dollars) or $1$ (expressed in Euros).
\end{corollary}
\begin{proof}
    Recall \eqref{E Qe}, which implies the identity $x_0 \Qe(R \leq T) = x_0 - E^{\Qd}[X_T]$.
\qed\end{proof}
The corresponding replicating strategy is the buy-and-hold strategy of one Euro.
\begin{corollary}[Put-call parity] \label{C pc parity}
    The prices of puts and calls simplify under the pricing operator $p^{\$}$ to
    \begin{align}
        p^{\$}(D^{P,\$}_K) &= E^{\Qd}[(K-X_T)^+]; \nonumber \\
        p^{\$}(D^{C,\$}_K) &= E^{\Qd}[(X_T-K)^+] + x_0 \Qe(R \leq T);  \label{E call}
    \end{align}
    moreover, the put-call parity
    \begin{align} \label{E PutCall}
        p^{\$}(D^{C,\$}_K) + K = p^{\$}(D^{P,\$}_K) + x_0
    \end{align}
    holds, where $K \in \R$ denotes the strike of the call $D^{C,\$}_K$ and put $D^{P,\$}_K$.
\end{corollary}
\begin{proof}
    The statement follows directly from \eqref{E p} and the linearity of expectation.
\qed\end{proof}
We refer to Madan and Yor \cite{MadanYor_Ito} for alternative representations
of the call price in \eqref{E call}.

Giddy \cite{Giddy} introduces the notion of \emph{international put-call equivalence} which
relates the price of a call in one currency with the price of a put in the other currency;
see also Grabbe \cite{Grabbe}.
\begin{corollary}[International put-call equivalence]
    The pricing operators $p^{\$}$ and $p^{\eu}$ satisfy international put-call equivalence:
    \begin{align*}
        p^{\$}\left(D^{C,\$}_K\right) &= x_0 K p^{\eu}\left(D^{P,\eu}_{\frac{1}{K}}\right);\\
        p^{\$}\left(D^{P,\$}_K\right) &= x_0 K p^{\eu}\left(D^{C,\eu}_{\frac{1}{K}}\right)
    \end{align*}
    for all $K > 0$.
\end{corollary}
\begin{proof}
    We obtain
    \begin{align*}
        x_0 K p^{\eu}\left(D^{P,\eu}_{\frac{1}{K}}\right) &=
            x_0 K \left(\E^{\Qe}\left[\left(\frac{1}{K} - Y_T\right)^+
                \1_{\{R > T \}}\right] + \E^{\Qe}\left[\frac{1}{K}\1_{\{R \leq T \}}\right]
                \right)\\
            &= x_0 \left(\E^{\Qe}\left[\left(\left(X_T - K\right)^+ \1_{\{R>T \}}\right) Y_T
                \right] + {\Qe}\left(R \leq T \right)
                \right)\\
            &= E^{\Qd}\left[(X_T-K)^+ \1_{\{S>T \}}\right] + x_0 \Qe(R \leq T)\\
            &= p^{\$}\left(D^{C,\$}_K\right),
    \end{align*}
    where we have used the identities of Corollary~\ref{C pc parity} and \eqref{E QdQe}. The second equivalence follows in the same
    way or from the put-call parity for model prices in \eqref{E PutCall}.
\qed\end{proof}

The next remark discusses how our result motivates and generalizes Lewis' Generalized Pricing Formulas.
\begin{remark}[Lewis' Generalized Pricing Formulas]
    Within Markovian stochastic volatiliy models, Lewis \cite{Lewis} derives call and put prices which exactly
    correspond to \eqref{E p} when applied to the call payoff $D^{C,\$}_K$ or put payoff
    $D^{P,\$}_K$.  Lewis starts from the postulate that
    put-call-parity holds and then shows that the correction term that is added to the expected payoff under $\Qd$
    corresponds to the probability of some process exploding under another measure (corresponding here to $\Qe$).
    We here start from an economic argument by defining the price as the minimal superreplicating cost for a contingent claim    under two, possibly non-equivalent measures that arise from a change of num\'eraire. We then show that this
    directly implies put-call parity for model prices.
    This approach also yields a generalization of
     Lewis' pricing formula to arbitrary, possibly path-dependent
    contingent claims. \qed
\end{remark}

\begin{example}[Singular measures (continued)]  \label{example singularII}
    We continue here our discussion of Example~\ref{example singularI} with $\Qd = \Q$ and
    $\Qe = \widehat{\Q}$.
    Although the exchange rate $X$ is a $\Qd$-local martingale,
    from the classical point of view of a Dollar investor
    the minimal superreplicating price of one Euro at time $T$
    is zero because under $\Qd$ there are only paths under which this contingent claim becomes worthless.
    However, by means of the correction term, \eqref{E p} yields a price $p^{\$}((X_T,1)) = x_0$, when
    considering the minimal joint $\Qd$- and $\Qe$-superreplicating price
    of one Euro.
    For the self-quantoed call $D^{SQC,\$}_K$, the classical price would be again zero;
    however, considering also the paths that the Euro investor under $\Qe$ can see,
    Theorem~\ref{T minimal} suggests a price $p^{\$}(D^{SQC,\$}_K)=\infty$ since $\E^{\Qe}[(X_T - K)^+] =
    \infty$. \qed
\end{example}
In many applications, however, the measures $\Qd$ and $\Qe$ do not become singular. Often, one measure is absolutely
continuous with respect to the other measure. In this case, the formulas for computing $p^{\$}(D)$ and $p^{\eu}(D)$ simplify:
\begin{corollary}[Absolutely continuous measures] \label{C absolutely}
    If $\Qd(S>T) = 1$, that is, if $Y$ is $\Qe$-martingale,, then $p^{\eu}$ can be computed as
    \begin{align*}
          p^{\eu}((D^{\$},D^{\eu})) = \E^{\Qe}[D^{\eu}].
    \end{align*}
    If $\Qe(R>T) = 1$, that is, if $X$ is $\Qd$-martingale, then
    \begin{align*}
        p^{\$}((D^{\$},D^{\eu})) = \E^{\Qd}[D^{\$}].
    \end{align*}
\end{corollary}
\begin{proof}
    Assume that $\Qd(S>T) = 1$. Then, Remark~\ref{R duality} implies that $\Qd$ is absolutely continuous with
    respect to $\Qe$. Thus, if a trading strategy superreplicates $D^{\eu}$ $\Qe$-almost surely for an
    Euro investor, then it also superreplicates $D^{\$}$ $\Qd$-almost surely for a
    Dollar investor. The second statement can be shown analogously.
\qed\end{proof}
\begin{example}[Reciprocal of the three-dimensional Bessel process] \label{example Bessel I}
    We set $ d = T = 1$ and let $X$ denote a nonnegative $\Qd$-local martingale
    identically distributed as the reciprocal of a three-dimensional Bessel process
    starting in $1$; in particular, there exists a Brownian motion
    $W = \{W_t\}\tInd$ such that
    \begin{align*}
        X_t = 1+\int_0^{t} X^2_u \dd W_u
    \end{align*}
    for all $t \in [0,T]$.
    It is well-known that $X$ is strictly positive and that $Y$ is a $\Qe$-Brownian motion stopped when it hits zero; see for example Delbaen and Schachermayer \cite{DS_Bessel}.
    Since $X$ is strictly positive, the discussion in
    Remark~\ref{R duality} yields that $\Qd$ is absolutely continuous with respect to $\Qe$;
    thus, Corollary~\ref{C absolutely} applies.

    Let us study the self-quantoed call $D^{SQC,\$}_K$. Since Brownian motion hits
    $0$ in any time interval with positive probability we obtain that $X$ hits $\infty$
    with positive $\Qe$-probability. This yields directly a
    minimal joint $\Qd$- and $\Qe$-superreplicating price $p^{\$}(D^{SQC,\$}_K) =\infty$.
    It is interesting to note that, as in Example~\ref{example singularII}, the classical
    price is finite:
    \begin{align*}
        E^{\Qd}[X_T (X_T - K)^+] &\leq E^{\Qd}[X_T^2] = \E^{\Qe}[X_T \1_{\{R > T\}}] \\
            &= \frac{1}{\sqrt{2 \pi T}} \int_0^\infty \frac{1}{y} \left(\exp\left(-\frac{(y-1)^2}{2T}\right)
            - \exp\left(-\frac{(y+1)^2}{2T}\right)\right) \dd y \\
            &< \infty
    \end{align*}
    for all $K \geq 0$, where we have plugged in the density of killed Brownian motion;
    see Exercise~III.1.15 in Revuz and Yor \cite{RY}.  \qed
\end{example}

We remark that, as a corollary of Remark~\ref{R duality}, in our setup there are only
positive ``bubbles'' under the corresponding measure.  A bubble is usually defined as the
difference of the current price and the expectation of the terminal value of an asset; that is,
$x_0 - \E^{\Qd}[X_T]$ and $1/x_0 - \E^{\Qe}[Y_T]$, respectively. It is possible, that
both bubbles are strictly positive; however, negative bubbles cannot occur by the
supermartingale property of the asset price processes under the corresponding measure.
This contrasts Jarrow and Protter \cite{JP_foreign}, where negative bubbles are discussed, however only when
considering the Dollar measure $\Qd$, which is not the  risk-neutral  measure of a Euro investor.

In the next section, we provide an interpretation of a bubble (lack of the martingale property of the exchange rate) as the possibility of a
hyperinflation under some dominating ``real-world'' measure $\Prob$.  If for both currencies such hyperinflations
have positive $\Prob$-probability, then $X$ and $Y$ both have a positive bubble.
\section{A physical measure}   \label{SS physical}
In this section, we start
by specifying a physical probability measure $\Prob$ instead of a risk-neutral probability measure $\Qd$.
We again assume a time horizon $T \in (0,\infty]$ and a filtered probability space $(\Omega, \CF_T, \{\CF_t\}\tInd, \Prob)$ that satisfies the technical conditions of Appendix~\ref{A conditions}. Let $X = \{X_t\}\tInd$ denote a process taking values in $[0,\infty]$ with right-continuous paths for all $\omega \in \Omega$. Define the stopping times $\{R_i\}_{i \in \N}, \{S_i\}_{i \in \N}, S$, and $R$ as in Section~\ref{S changemeasure} and assume that $\Prob(S_i \wedge T< S) = 1$ for all $i \in \N$,  $X_{(R+t) \wedge T} = \infty$ if $R \leq T$ and  $X_{(S+t) \wedge T} = 0$ if $S \leq T$ for all $t \geq 0$. In particular, this assumption implies that no oscillations can occur; that is, the events $H^{\$} :=\{R \leq T\}$ and $H^{\eu} :=\{S \leq T\}$ are disjoint. Suppose that $\Prob(H^{\$}) <1$ and $\Prob(H^{\eu}) <1$. Define again $Y = \{Y_t\}\tInd$ by $Y_t := 1/X_t \1_{\{R > t\}}$ for all $t \in [0,T]$.

Under the physical probability measure $\Prob$  the events $H^{\$}$ and $H^{\eu}$ may both have positive probability. We interpret these events as 
the complete devaluation (hyperinflation) of the Dollar or Euro currency with respect to the other. Such hyperinflations have been observed; for example, the exchange rate between the American and German currencies changed by a factor of $10^{10}$ from January 1922 to December 1923, as described in Sargent \cite{Sargent_1982}.
During a hyperinflation, the interest rate of the inflating currency tends to become very large, and so far, we have assumed zero interest rates.
We may reinterpret $X_T$ as the forward exchange rate of Dollars per Euro at time $T$, as opposed to a spot exchange rate.  This is consistent with various interest rate assumptions, including the possibility that the Dollar (respectively Euro) interest rate should explode when the Dollar (respectively Euro) experiences a hyperinflation.

 If $\Prob(H^{\$}) > 0$ and $\Prob(H^{\eu}) > 0$, then no risk-neutral measure equivalent to $\Prob$ can exist
such that either $X$ or $Y$ follow local martingale dynamics. Nevertheless, pricing and hedging of contingent claims still might  be possible. Towards this end,
let us introduce the two artificial measures
\begin{align*}
    \Prob^{\$}(\cdot) &:= \Prob(\cdot| H^{\$^C}) = \Prob(\cdot| R>T);  \\
    \Prob^{\eu}(\cdot) &:= \Prob(\cdot| H^{\eu^C}) = \Prob(\cdot| S>T), 
\end{align*}
where we have conditioned the physical measure $\Prob$ on the events $H^{\$^C}$ and $H^{\eu^C}$
that no hyperinflation occurs. 
Note that both measures $\Prob^{\$}$ and $\Prob^{\eu}$ are absolutely continuous with
respect to $\Prob$ and that $\Prob$ is absolutely continuous with respect to their average
$(\Prob^{\$}+ \Prob^{\eu})/2$. 

As in Section~\ref{S replication}, for some $d \in \N$, 
we assume the existence of $d+1$ tradable assets with nonnegative price processes, denoted by $S^{\$} = \{S^{\$, (i)}\}_{i = 0, 1, \ldots d}$ in Dollars and by $S^{\eu} =\{S^{\eu, (i)}\}_{i = 0, 1, \ldots d}$ in Euros, respectively; as before, with
$S^{\$, (i)} = \{S^{\$, (i)}_t\}\tInd$ and $S^{\eu, (i)} = \{S^{\eu, (i)}_t\}\tInd$ for all $i = 0, 1, \ldots, d$. We also assume that $S^{\eu, (i)}_t \1_{\{R \wedge S>t\}} = S^{\$, (i)}_t Y_t \1_{\{R \wedge S>t\}}$ for all $t \in [0,T]$ and $i = 0, 1, \ldots, d$ and that $S^{\$, (1)} = X$ and $S^{\eu, (0)} = Y$. Moreover, we assume that $S^{\$}$ and $S^{\eu}$ have \cadlag{} paths $\Prob^{\$}$ and $\Prob^{\eu}$-almost surely, respectively. 

Suppose that 
there exists exactly one probability measure $\Qd$ ($\Qe$) that is equivalent to 
        $\Prob^{\$}$ ($\Prob^{\eu}$) and under which the processes $S^{\$}$ ($S^{\eu}$) are local martingales.
Consider the condition of \emph{no obvious hyperinflations (NOH)}:
\begin{enumerate}
	\item[(NOH)]  The probability measures $\Prob^{\$}$ and $\Prob^{\eu}$ are equivalent on
        $\CF_{(R_i \wedge S_j)-}$ for all $i,j \in \N$.
\end{enumerate}
This condition corresponds to an environment in which, at no time, one knows that a certain hyperinflation will occur $\Prob$-almost surely; that is, hyperinflations occur as a surprise. To see this, assume that the condition (NOH) does not hold. Then there exist $i,j \in \N$ and a set $A \in \CF_{(R_i \wedge S_j)-}$ such that, for example,  $\Prob^{\$}(A) = 0$ and  $\Prob^{\eu}(A) > 0$, which implies $A \subset H^{\$}$ (modulo $\Prob$). Thus, for some paths one knows that a hyperinflation will occur before it occurs.
As the next lemma shows, the condition (NOH)  brings us back to the framework of Theorems~\ref{T numeraire} and \ref{T minimal}:
\begin{lemma}[(NOH)  and change of num\'eraire]  \label{L consistency}
    The following two conditions are equivalent:
    \begin{enumerate}
    \item[(i)] The condition (NOH)  holds.
    \item[(ii)] The equality in \eqref{E Qe0} of Theorem~\ref{T numeraire} holds for all stopping times $\tau$ and $A \in \CF_{\tau \wedge T}$ with $\widehat{\Q}$ replaced by $\Qe$ and $\Q$ replaced by $\Qd$.
    \end{enumerate}
\end{lemma}
The proof of this equivalence is based on the assumption that $\Qe$ is the unique probability measure equivalent to $\Prob^{\eu}$ such that  $S^{\eu}$ are ${\Qe}$-local martingales. It is contained in Appendix~\ref{A proof 4}. 

The next proposition yields that the minimal cost (expressed in Dollars) for replicating a contingent claim $D = (D^{\$}, D^{\eu})$ $\Prob$-almost surely is given by \eqref{E p}  in Theorem~\ref{T minimal} if condition (NOH)  holds.  More precisely, one can find a trading strategy $\eta$, in the sense of Definition~\ref{D trading}, such that the corresponding terminal wealth satisfies $\Prob^{\$}(V^{\$,\eta}_T = D^{\$}) = 1= \Prob^{\eu}(V^{\eu,\eta}_T = D^{\eu})$. Since $\Prob$ and $(\Prob^{\$}+ \Prob^{\eu})/2$ are equivalent, we interpret $\eta$ as a replication strategy under the physical measure $\Prob$.
\begin{proposition}[Minimal replication cost under $\Prob$]  \label{P minimal P}
Assume that condition (NOH)  holds. Then  the minimal replicating cost for a contingent claim $D = (D^{\$}, D^{\eu})$ under $\Prob$ is exactly the
one computed in Theorem~\ref{T minimal}.  
\end{proposition}
\begin{proof}
	Theorem~\ref{T minimal} yields the minimal joint replicating cost of a claim under $\Qd$ and under any extension of the corresponding measure after a change of num\'eraire.  However, Lemma~\ref{L consistency}
	shows that $\Qe$ is exactly such an extension if (NOH)  holds. This yields the assertion since $\Qd$ and $\Qe$ are equivalent to $\Prob^{\$}$ and $\Prob^{\eu}$, respectively.
\qed\end{proof}

We also obtain an interpretation of the lack of martingale property of $X$ under the risk-neutral measure $\Qd$ as the possibility of an explosion under the physical measure $\Prob$:
\begin{corollary}[Interpretation of the lack of the martingale property]   \label{C interpretation}
    Assuming the condition (NOH), we have $\Prob(H^{\$}) > 0$ if
    and only if $X$ is a strict $\Qd$-local martingale; equivalently, we have 
    $\Prob(H^{\eu}) > 0$ if
    and only if $Y$ is a strict $\Qe$-local martingale.
\end{corollary}
\begin{proof}
    Note that $\Prob(H^{\$}) > 0$ if and only if $\Prob^{\eu}(H^{\$}) > 0$, which is equivalent to  $\Qe(H^{\$}) > 0$. Lemma~\ref{L consistency} and Remark~\ref{R duality} then yield the first equivalence of the assertion.
The second equivalence follows in the same manner.
\qed\end{proof}

If the condition (NOH)  does not hold then Theorem~\ref{T minimal} still provides an upper bound for the minimal replicating cost of a contingent claim. However, 
as the next example shows, the expression in \eqref{E p} usually does not give the smallest minimal replicating cost under $\Prob$.

\begin{example}[Condition (NOH)  not satisfied]
    We fix $d = 1$ and $T=1$. Let $U$ denote an $\CF_0$--measurable random variable, taking values in $\{-1,1\}$ with $\Prob(U = 1) \in (0,1)$.
    Furthermore, define $X = \{X_t\}\tInd$ by $X_t := (Z_t)^U$ for all $t \in [0,1]$, where $Z = \{Z_t\}\tInd$ has the same distribution as the process in \eqref{E singular X}.
Thus, $H^{\$} = \{U = -1\}$ and $H^{\eu} = \{U = 1\}$ and the condition (NOH)  is not satisfied.

	Consider the contingent claim $D = (1,1)$, which pays either one Dollar if the Dollar does not hyperinflate or otherwise one Euro. Then, \eqref{E p} yields the price (in Dollars) $p^{\$}(D) = 1 + 1 = 2$. However, at time zero, it is already well-known which of the two currencies defaults, as $U$ is $\CF_0$--measurable.  Thus, the trading strategy $\eta = (\1_{\{U = 1\}}, \1_{\{U = -1\}})$, holding one unit of the corresponding currency, perfectly replicates the contingent claim at an initial cost of only one Dollar.
\qed
\end{example}

\section{Conclusion}
Based on a replication argument, we introduced a novel pricing operator for contingent claims that restores put-call parity and international put-call equivalence for model prices. If the underlying is a true martingale, our pricing operator is just the classical replication price. Furthermore, we interpreted the lack of martingale property of an underlying price process under the risk-neutral probability as the positive probability of an explosion (hyperinflation) under some dominating physical measure.

Two directions of future research arise. First, we focused on the case of two currencies, corresponding to one exchange rate, only. It would be interesting to extend the results of this paper to multiple currencies and to find a consistent way to describe devaluations of currencies with respect to several other currencies. The num\'eraire-free approach taken in Yan \cite{Yan_new} might be very useful.
Second, throughout this paper we relied on the assumption that markets are complete. Again, it would be interesting to consider incomplete markets and to develop a theory of joint superreplication in such markets.

\begin{acknowledgements}
We thank Sara Biagini, Zhenyu Cui, Pierre Henry-Labord\`ere,  Ioannis Karatzas, Kostas Kardaras, Martin Klimmek, Alex Lipton, and Nicolas Perkowski for
their helpful comments on an early version of this paper. In particular, we are deeply indebted to
 Sergio Pulido for his careful reading of this paper and for several helpful discussions. We are grateful to two anonymous referees, an associate editor, and Martin Schweizer for very helpful suggestions, which substantially improved this paper.
\end{acknowledgements}

\appendix
\section{Local martingales on stochastic intervals}  \label{A stoch interval}
	In this appendix, we provide some technical results for stochastic processes that satisfy the local martingale property up to a stopping time.  Such stochastic processes appear throughout this paper.

	Similar to Perkowski and Ruf \cite{Perkowski_Ruf}, we consider the time set $\mathcal{T}:= [0,\infty] \bigcup \{\mathfrak{T}\}$, where $\mathfrak{T}$ represents a time ``beyond horizon;'' the natural ordering is  extended to $\mathcal{T}$ by $t < \mathfrak{T}$ for all $t \in [0,\infty]$.   For any $t \in \mathcal{T}$ and for any sequence $\{t_i\}_{i \in \N}$ with $t_i \in \mathcal{T}$ for all $i \in \N$ we write $\lim_{i \uparrow \infty} t_i = t$ if either (a) $t = \mathfrak{T}$ and 
	$\inf_{i \geq j} \{t_i\}  = \mathfrak{T}$ for some $j \in \N$ or if (b) $t < \mathfrak{T}$, $\sup_{i \geq j} \{t_i\} < \mathfrak{T}$, and $\lim_{i \uparrow \infty; i \geq j} t_i = t$ for some $j \in \N$.

	Throughout this appendix, we fix a time horizon $T \in (0,\infty]$, an arbitrary stochastic basis $(\Omega, \CF_T, \{\CF_t\}_{t \in [0, T]}, \Prob)$, and a process $N = \{N_t\}_{t \in [0,T]}$ taking values in $[-\infty, \infty]$. For a $\mathcal{T}$-valued random variable $\tau$ we define the stochastic process $N^\tau = \{N_t^\tau\}_{t \in [0,T]} :=\{N_{t \wedge \tau}\}_{t \in [0,T]}$.
	Throughout this appendix, we shall fix a stopping time $\tau$, which is a map $\tau: \Omega \rightarrow \mathcal{T}$ such that $\{\tau \leq t\} \in \CF_t$ for all $t \in [0,T]$.
	If not specified further, all (in)equalities are interpreted in the $\Prob$-almost sure sense.	
	
We start with a definition:
\begin{definition}[Local martingale on stochastic interval] \label{D localS}
	We call $N$ 
	\begin{enumerate}
		\item[(1)] a local martingale on $[0, \tau]$ if there exists a  non-decreasing sequence of stopping times $\{\tau_i\}_{i \in \N}$
	with $\lim_{i \uparrow \infty} \tau_i > \tau \wedge T$  such that $N^{\tau_i \wedge \tau}$ is a martingale for all $i \in \N$;
	\item[(2)] a local martingale on $[0, \tau)$ if there exists a  non-decreasing sequence of stopping times $\{\tau_i\}_{i \in \N}$ with $\lim_{i \uparrow \infty} \tau_i = \tau$ such that
    $N^{\tau_i}$ is a martingale for all $i \in \N$.
 \qed
 \end{enumerate}
\end{definition}
In particular, if $T = \tau = \infty$, then a local martingale on $[0,\tau)$ corresponds exactly to the usual notion of a local martingale.  Observe that if $N$ is a local martingale on $[0,\tau]$ then it is a local martingale on $[0,\tau)$.  If the definition of local martingale on $[0,\tau)$ required additional the assumption that $\tau_i < \tau$ for all $i \in \N$ (something that Definition~\ref{D localS} does not require), this implication would in general not hold true; consider for example a compensated Poisson process and $\tau$ the time of its first jump. Observe also that if $\widetilde{\tau}$ is a stopping time with $\widetilde{\tau} \wedge T < \epsilon \vee \tau$ for all $\epsilon > 0$ then any local martingale on $[0,\tau)$ is also a local martingale on $[0,\widetilde{\tau}]$.

In the following, we repeatedly will use the fact that $$N^{{\eta_1} \vee \eta_2} = N^{{\eta_1} } + N^{ \eta_2} - N^{{\eta_1} \wedge \eta_2}$$ is a (local) martingale if $N^{{\eta_1} }$ and $N^{ \eta_2}$ are (local) martingales for some stopping times ${\eta_1}$ and $\eta_2$. The next lemma is useful in  several of the proofs in this paper:
\begin{lemma}[Localization sequence for a local martingale on a stochastic interval]  \label{L locsequence}
The following two statements are equivalent:
		\begin{enumerate}
			\item[(a.1)] $N$ is a local martingale on $[0, \tau]$. 
			\item[(a.2)] There exists a non-decreasing sequence of stopping times $\{\tau_i\}_{i \in \N}$ such that
	 $\lim_{i \uparrow \infty} \tau_i > \tau \wedge T$ and that $N$ is a local martingale on $[0,{\tau_i \wedge \tau}]$ for all $i \in \N$.
		\end{enumerate}
The following three statements are equivalent:
		\begin{enumerate}
			\item[(b.1)] $N$ is a local martingale on $[0, \tau)$.
			\item[(b.2)] There exists a non-decreasing sequence of  stopping times $\{\tau_i\}_{i \in \N}$
	such that $\lim_{i \uparrow \infty} \tau_i = \tau$ and that $N$ is a local martingale on $[0,\tau_i]$ for all $i \in \N$.
			\item[(b.3)] There exists a non-decreasing sequence of  stopping times $\{\tau_i\}_{i \in \N}$
	such that $\lim_{i \uparrow \infty} \tau_i = \tau$ and that $N$ is a local martingale on $[0, \tau_i)$ for all $i \in \N$.
		\end{enumerate}
\end{lemma}
\begin{proof}
	For the first part, we only need to show the implication from (a.2) to (a.1). 
Thus, assume (a.2), which yields that there exists a stopping time ${\eta}_i$ with $\Prob({\eta}_i \leq \tau_i \wedge \tau \wedge T) \leq 2^{-i}$ such that $N^{{\eta}_i \wedge {\tau}_i  \wedge \tau}$ is a  martingale for all $i \in \N$.  Define $\widetilde{\tau}_i =  \max_{j \in \{1, \ldots, i\}} \{{\eta}_j \wedge {\tau}_j\}$ for all $i \in \N$ and observe that $N^{\widetilde{\tau}_i \wedge \tau}$ is a martingale for all $i \in \N$ and that $\lim_{i \uparrow \infty} \widetilde{\tau}_i > \tau \wedge T$. This shows that (a.1) holds.  

	For the second part, we only need to show the implication from (b.3) to (b.1). 
Assume now (b.3). Then there exists a non-decreasing sequence of  stopping times  $\{{\eta}_i\}_{i \in \N}$ such that 
$$\Prob\left(\left(\{\tau_i = \mathfrak{T}\} \bigcap \{\eta_i < \mathfrak{T}\} \right)
\bigcup \left(\{\tau_i < \mathfrak{T}\} \bigcap \{\eta_i <  (\tau_i - 2^{-i}) \wedge i\}\right) \right) \leq 2^{-i}$$ and $N^{\eta_i}$ is a martingale for all $i \in \N$.  
Define $\widetilde{\tau}_i := \max_{j \in \{1, \ldots, i\}} \{\eta_i\} \wedge \tau$ and observe that $N^{\widetilde{\tau}_i}$ is a martingale for all $i \in \N$ and that $\lim_{i \uparrow \infty} \widetilde{\tau}_i = \tau$, which yields (b.1).
\qed\end{proof}

For the next lemma, observe that the random times
\begin{align}  \label{E rhoj}
\rho_j := \inf \left\{t \in [0,T] | N_t^\tau >  j\right\}
\end{align}
with $\inf \emptyset := \mathfrak{T}$ for all $j \in \N$   take values in $[0,\tau \wedge T] \bigcup \mathfrak{T}$ and are stopping times if 
the underlying filtration $\{\CF_t\}_{t \in [0,T]}$ is right-continuous and $N(\omega)$ is a right-continuous path for all $\omega \in \Omega$; see, for example, Problem~1.2.6 in Karatzas and Shreve \cite{KS1}. 

\begin{lemma}[Localization sequence for nonnegative local martingale]  \label{L locsequence2}
	Assume that the underlying filtration $\{\CF_t\}_{t \in [0,T]}$ is right-continuous and $N(\omega)$ is a right-continuous path taking values in $[0,\infty]$ for all $\omega \in \Omega$.  Define the stopping times $\{\rho_j\}_{j \in \N}$ as in \eqref{E rhoj} and $\rho := \lim_{j \uparrow \infty} \rho_j$. 
Then the following statements hold:
	\begin{enumerate}
		\item[(i)] If $N^{\rho_j \wedge \tau}$ is a supermartingale for all $j \in \N$ (in particular, if $N^\tau$ is a supermartingale)  then $\rho= \mathfrak{T}$.
		\item[(ii)] If $N^{\rho_j \wedge \tau_i^{(j)}}$ is a supermartingale for all $i,j \in \N$ for some non-decreasing sequences of  stopping times $\{\tau_i^{(j)}\}_{i \in \N}$
	with $\lim_{i \uparrow \infty} \tau_i^{(j)} = \tau$ for all $j \in \N$ then $\rho \geq \tau$.
	\end{enumerate}	
	The following statements are equivalent:
			\begin{enumerate}
				\item[(a.1)] $N$ is a local martingale on $[0,\tau]$; 
				\item[(a.2)] $N^{\rho_j \wedge \tau}$ is a uniformly integrable martingale for all $j \in \N$;
				\item[(a.3)]  $N^{\rho_j}$ is a local martingale on $[0,\tau]$ for all $j \in \N$.
			\end{enumerate}
	The following statements are equivalent:
			\begin{enumerate}
				\item[(b.1)] $N$ is a local martingale on $[0,\tau)$; 
				\item[(b.2)] $N^{\rho_j \wedge \tau_i}$ is a uniformly integrable martingale  for all $i,j \in \N$ for some non-decreasing sequence of  stopping times $\{\tau_i\}_{i \in \N}$
	with $\lim_{i \uparrow \infty} \tau_i = \tau$;
				\item[(b.3)] $N^{\rho_j}$ is a local martingale on $[0,\tau)$ for all $j \in \N$.
			\end{enumerate}
\end{lemma}
\begin{proof}
	Assume that $N^{\rho_j \wedge \tau}$ is a nonnegative supermartingale  and observe that $N^{\rho_j \wedge \tau}_T \geq j$ if $\rho_j \leq \tau \wedge T$; thus, $\Prob(\rho_j \leq \tau \wedge T) \leq N_0 / j$ for all $j \in \N$, which yields (i).
Next, assume that  there exist sequences of stopping times $\{\tau_i^{(j)}\}_{i \in \N}$ such that  $N^{\rho_j \wedge \tau_i^{(j)}}$ is a supermartingale for all $i,j \in \N$.  Fix a sequence $\{i_j\}_{j \in N}$ so that
$$\Prob\left(\left(\{\tau = \mathfrak{T}\} \bigcap \{\tau_{i_j}^{(j)} < \mathfrak{T}\} \right)
\bigcup \left(\{\tau < \mathfrak{T}\} \bigcap \{\tau_{i_j}^{(j)} <  (\tau - 2^{-j}) \wedge j\}\right) \right) \leq 2^{-j}.$$
Then we have  $N_0 \geq \E[ N^{\rho_j \wedge \tau_{i_j}^{(j)}}_T] \geq j \Prob(\rho_j \leq \tau_{i_j}^{(j)} \wedge T)$ and thus  
$$\Prob\left(\left(\{\tau = \mathfrak{T}\} \bigcap \{\rho_j < \mathfrak{T}\} \right)
\bigcup \left(\{\tau < \mathfrak{T}\} \bigcap \{\rho_j <  (\tau - 2^{-j}) \wedge j\}\right) \right) \leq \frac{N_0}{j} + 2^{-j}$$
for all $j \in \N$, which yields (ii).

Now, assume (a.1) and observe that $\sup_{t \in [0,T]} \{N_t^{\rho_j \wedge \tau}\} \leq  j + N_T^{\rho_j \wedge \tau}$ and that $N_T^{\rho_j \wedge \tau}$ is integrable  for all $j\in \N$ since $N^\tau$ is a supermartingale. This observation in conjunction with dominated convergence shows (a.2). Next, assume (a.3) and observe that  $N^{\rho_j}$ is a supermartingale on $[0, \tau]$ for all $j \in \N$, and thus, $\rho=\mathfrak{T}$ by (i). The first part of Lemma~\ref{L locsequence} then yields (a.1).

Now, assume (b.1), which gives the existence of a non-decreasing sequence of  stopping times $\{\tau_i\}_{i \in \N}$ with $\lim_{i \uparrow \infty} \tau_i = \tau$ such that $N$ is a local martingale on $[0,\tau_i]$ for all $i \in \N$. Using the implication of (a.1) to (a.2) with $\tau$ replaced by $\tau_i$ for all $i \in \N$, we observe that (b.2) holds. Next, assume (b.3). Then (ii) yields that $\rho \geq \tau$ and the second part of Lemma~\ref{L locsequence} then yields (b.1).
 \qed
\end{proof}  

Note that the implication of (b.3) to (b.1) in Lemma~\ref{L locsequence2} with $T=\tau= \infty$ yields that any nonnegative right-continuous process $N$ is automatically a local martingale (on $[0,\infty)$) if $N^{\rho_j}$ is a local martingale (on $[0,\infty)$)  for all $j \in \N$. Furthermore, by (ii), $N^{\rho_j}$ is a supermartingale for all $j \in \N$ if and only if $N$ is a supermartingale.

We call the stopping time $\tau$ \emph{foretellable} if there exists a non-decreasing sequence of  stopping times  $\{\tau_i\}_{i \in \N}$ such that $\lim_{i \uparrow \infty} \tau_i = \tau$ (in particular, there exists some $i(\omega) \in \N$ with $\tau_{i(\omega)}(\omega) = \mathfrak{T}$ if $\tau(\omega) = \mathfrak{T})$ and $\tau_i \wedge T < \tau \vee \epsilon$ for all $i \in \N$ and $\epsilon>0$. We then call $\{\tau_i\}_{i \in \N}$ an \emph{announcing} sequence of $\tau$.

The following result illustrates that a nonnegative
local martingale on a half-open
stochastic interval (with respect to a foretellable stopping time) can be extended to one on a closed interval. For example, if $N$ is defined by $N_t := \1_{\{t < \tau\}}$ for all $t \in [0,T]$, then $N$ can be extended to a process $M = \{M_t\}\tInd$ with $M_t := 1$ for all $t \in [0,T]$, representing a local martingale on $[0, T]$.
\begin{proposition}[Extension of local martingales on a stochastic interval] \label{P extension}
Suppose that the assumptions of Lemma~\ref{L locsequence2} hold and    
assume that $\tau$ is foretellable and that $N$ is a local martingale on $[0,\tau)$.    Then, there exists a unique local martingale $M = \{M_t\}_{t \in [0,T]}$ on $[0,T]$ such that we have $M = M^\tau$, $\{M_t \1_{\{t <\tau\}}\}_{t \in [0,T]} = \{N_t \1_{\{t <\tau\}}\}_{t \in [0,T]}$, $M_0 = N_0$, and, moreover,  $\lim_{s \uparrow \tau(\omega)} M_s(\omega) = M_t(\omega)$ for all $\omega \in \Omega$ with $\tau(\omega) \notin \{0,\mathfrak{T}\}$.
    The process $M$ has nonnegative and right-continuous paths.
\end{proposition}
\begin{proof}
	The uniqueness of $M$ follows directly from its left-continuity at time $\tau$.  Let $\{\tau_i\}_{i \in \N}$ denote an announcing sequence of $\tau$ and let $\{\widetilde{\tau}_i\}_{i \in \N}$  denote a non-decreasing sequence of stopping times such that $N^{\widetilde{\tau}_i}$ is a martingale for all $i \in \N$ and $\lim_{i \uparrow \infty} \widetilde{\tau}_i = \tau$. We assume, without loss of generality, that $\tau_i = \tau_i \wedge \widetilde{\tau}_i$ for all $i \in \N$. Observe that $N^{\tau_i}$ is a nonnegative supermartingale for all $i \in \N$. By imitating the proof of Theorem~1.3.15 in Karatzas and Shreve \cite{KS1} based on Doob's up- and downcrossing inequalities (replace therein $n$ by $\tau_n$ for all $n \in \N$ and $\infty$ by $\tau$) we obtain that  $M_t := \lim_{i \uparrow \infty}  N_t^{\tau_i}$ for all $t \in [0,T]$ exists.

	We need to show that $M$, defined in this way, is a local martingale on $[0,T]$. By Lemma~\ref{L locsequence2}, it is sufficient to show that $M^{\widetilde{\rho}_j}$ is a martingale for all $j\in \N$, where $\widetilde{\rho}_j := \inf \{t \in [0,T] | {M}_t > j\}$ 
 with $\inf \emptyset := \mathfrak{T}$.  Fix an arbitrary $j \in \N$ and observe that, 
	 by dominated and monotone convergence,
	  \begin{align*}
	  	\E\left[M_T^{\widetilde{\rho}_j}\right] &=  \E\left[\lim_{i \uparrow \infty}N_T^{\tau_i}  \1_{\{\tau_i<  \widetilde{\rho}_j\}}\right]  + \E\left[ \lim_{i \uparrow \infty} N_T^{\widetilde{\rho}_j}  \1_{\{\tau_i \geq \widetilde{\rho}_j\}}\right] = \lim_{i \uparrow \infty} \E\left[N_T^{{\widetilde{\rho}}_j \wedge \tau_i}\right] \\&=  N_0 = M_0,
	\end{align*}
	 which yields the statement since, by Fatou's lemma, $M^{\widetilde{\rho}_j}$ is a supermartingale.
\qed\end{proof}

We warn the reader that usually $M_T^\tau \neq N^\tau_T$, even if $N$ is a martingale on $[0,\tau]$ since $N$ needs not be left-continuous at $\tau$.
We also refer the reader  to the related Exercise~IV.1.48 in Revuz and Yor \cite{RY}, where the case of not
necessarily nonnegative local martingales is treated.

\section{Conditions on the filtration in Sections~\ref{S changemeasure}, \ref{S replication}, and \ref{SS physical}}  \label{A conditions}
In this appendix, we discuss the technical assumptions on the underlying filtration that are necessary for the results in Sections~\ref{S changemeasure}, \ref{S replication}, and \ref{SS physical}.
Throughout this appendix, we fix a time horizon $T \in (0,\infty]$ and denote a set of states by $\Omega \not= \emptyset$ and a filtration by $\{{\CF}_t\}_{t \in [0,T]}$.   

We refer to Appendix~\ref{A stoch interval} for the definition of a stopping time. 
For any stopping time $\tau$, we define $${\CF}_{\tau} := \left\{A \in {\CF}_T| A \cap \{\tau \leq t\} \in {\CF}_t \text{ for all } t  \in [0,T]\right\}$$ and $${\CF}_{\tau-} :=  \sigma\left(\left\{A \cap \{\tau > t\}| A \in {\CF}_t \text{ for some } t \in [0,T]\right\} \bigcup {\CF}_0^0\right)$$ 
if $\{{\CF}_t\}_{t \in [0,T]}$ is the right-continuous modification of a filtration $\{{\CF}_t^0\}_{t \in [0,T]}$;
see page~156 in F\"ollmer \cite{F1972}.

In Section~\ref{S changemeasure}, we are constructing a probability measure on $(\Omega, \CF_{R-})$ for a certain stopping time $R := \lim_{i \uparrow \infty} R_i$, where $\{R_i\}_{i \in \N}$ is a sequence of nondecreasing stopping times, defined in  Section~\ref{S changemeasure}.
This construction is based on an extension theorem; more precisely, on Theorem~V.4.1 in Parthasarathy \cite{Pa}, and thus, requires certain technical assumptions  on the  filtration $\{{\CF}\}_{t \in [0,T]}$. Specifically, we shall require in Sections~\ref{S changemeasure}, \ref{S replication}, and \ref{SS physical} that 
\begin{itemize}
	\item[(i)] $\{{\CF}_t\}_{t \in [0,T]}$ is the right-continuous modification of a filtration $\{{\CF}_t^0\}_{t \in [0,T]}$ and
	\item[(ii)]  $\{{\CF}_{R_i-}\}_{i \in \N}$ is a standard system, as defined in Section~6 of F\"ollmer \cite{F1972}.
\end{itemize} 
Furthermore, in Section~\ref{S replication}, we shall require that
\begin{itemize}
	\item[(iii)] any probability measure $\Prob$ on $(\Omega, \CF_{R-})$ can be extended to a probability measure $\widetilde{\Prob}$ on  $(\Omega, \CF_T)$.
\end{itemize} 

A sufficient condition for requirement (ii) is that 
$$\{\widehat{\CF}_{t}\}_{t \in [0,T]} :=  \{{\CF}_t \cap {\CF}_{R_-}\}_{t \in [0,T]}$$ is the right-continuous modification of a \emph{standard system} (\emph{RCMSS}); see Remark~6.1.1 in F\"ollmer \cite{F1972}, applied to the filtration $\{\mathcal{G}_t\}_{t \geq 0}$ with $\mathcal{G}_t := {\widehat{\CF}}_{1/(1-t) - 1}$, if $T=\infty$, and $\mathcal{G}_t := {\widehat{\CF}}_{t T},$ otherwise, for all $t \in [0,1]$ and $\mathcal{G}_t = {\widehat{\CF}}_T$ for all $t > 1$.
 We remark that $\{\widehat{\CF}_{t}\}_{t \in [0,T]}$ then does usually not satisfy the ``usual conditions'' as it is not completed under some probability measure.
Observe that if  $\{{\CF}_t^0\}_{t \in [0,T]}$ is a standard system then so is $\{{\CF}^0_t \cap {\CF}_{R_-}\}_{t \in [0,T]}$.

In the following, we provide a canonical example for $\Omega$ and for a filtration $\{{\CF}_t\}_{t \in [0,T]}$, such that $\{{\CF}_{t} \cap {\CF}_{R-}\}_{t \in [0,T]}$ is RCMSS.  This example provides a sufficiently rich structure so that one might as well assume, throughout this paper, that the underlying filtered measurable space is of that form.

 Towards this end, let $E$ denote any locally compact space with countable base (for instance, $E=\R^n$ for some $n \in \N$) and let $\Omega$ denote the space of right-continuous paths $\omega: [0,T] \rightarrow [0,\infty] \times E$ whose first component $\omega^{(1)}$ of $\omega$ satisfies $\omega^{(1)}(R(\omega) + t) = \infty$ for all $t \geq 0$, and that have left limits on $(0,R(\omega))$, where $R(\omega)$ denotes the first time that $\omega^{(1)} = \infty$.  Let $\{{\CF}_t^0\}_{t \in [0,T]}$ denote the filtration generated by the paths and $\{\CF_t\}\tInd$ its right-continuous modification. Then it follows, as in Dellacherie \cite{Dellacherie1969}, Meyer \cite{M}, and Example~6.3.2 of F\"ollmer \cite{F1972}, that $\{{\CF}_{t} \cap {\CF}_{R-}\}_{t \in [0,T]}$ is RCMSS. We identify the process $X(\omega)$, which appears in Section~\ref{S changemeasure}, with the first coordinate of $\omega$.

Observe that  in the canonical setup of the last paragraph, the extension of requirement (iii) always exists. To see this, define  $\widetilde{\Prob}(A) := \Prob(\omega^{R-} \in  A)$ for all  $A \in \CF_T$, where  $\omega^{R-} \in \Omega$ is given, for all $\omega \in \Omega$, by $$\omega^{R-}(t) := \omega(t) \1_{t < R(\omega)} + (\infty \times e) \1_{t \geq R(\omega)}$$ for some $e \in E$ for all $t \in [0,T]$. This specific construction then yields one extension $\widetilde{\Prob}$ on  $(\Omega, \CF_T)$.

\section{Proof of Proposition~\ref{P Equivalence} and further statements concerning the change of measure in Section~\ref{S changemeasure}}  \label{A proofs}
In this appendix, we provide additional statements on the change of measure suggested in Section~\ref{S changemeasure} and the proof of Proposition~\ref{P Equivalence}. We refer to Appendix~\ref{A stoch interval} for the definition of a stopping time.

Below, we shall rely on the next lemma:
\begin{lemma}[Convergence of stopping times] \label{L stoppingtimes}
	Assume the setup of Theorem~\ref{T numeraire} and fix a stopping time $\tau$.
       Then we have 
        $\Q(S>\tau) = 0$ if and only if $\widehat{\Q}(R>\tau) = 0$.
\end{lemma}
\begin{proof}
	Without loss of generality, we set $x_0 = 1$. Then \eqref{E Qe0} yields that 
	$$\widehat{\Q}(R>\tau) = \widehat{\Q} \left(\{\tau \leq T\} \cap {\{R > \tau \wedge T\}}\right) = \E^{{\Q}} \left[\1_{\{\tau \leq T\}} X^\tau_T\right]  \leq \E^{{\Q}} \left[\1_{\{S > \tau \}} X^\tau_T\right],$$
	which yields one direction of the statement. The other direction follows from \eqref{E QdQe} in the same manner.
\qed\end{proof}

Next, we formulate a generalized version of Bayes' formula.  If $X$ is a $\Q$-martingale, this formula has been well-known;
see for example Lemma~3.5.3 in Karatzas and Shreve \cite{KS1}. If $X$ is a strictly positive continuous
$\Q$-local martingale,  Bayes' formula has been derived in Ruf \cite{Ruf_ots}.
\begin{proposition}[Bayes' formula]  \label{P Bayes}
	Assume the setup of Theorem~\ref{T numeraire} and fix two stopping times $\rho, \tau$ with $\rho \leq \tau $ $\Q$- and $\widehat{\Q}$-almost surely and a  $\CF_{\tau \wedge T}$--measurable random variables $Z \in [0,\infty]$. Then
    we have the Bayes' formula
    \begin{align}
            \E^{\widehat{\Q}} \left[\left.\left(Z \1_{\{R > \tau \wedge T\}}\right) Y^\tau_T\right|\CF_{\rho}\right] \1_{\{S > \rho\wedge T\}}
                &= \E^{\Q} \left[\left.Z \1_{\{S > \tau\wedge T\}} \right|\CF_{\rho}\right] \1_{\{R>\rho\wedge T\}} Y^\rho_T   \label{E Bayes1}\\
                &\left(= \E^{\Q} \left[\left.Z \1_{\{S>\tau\wedge T\}} \right|\CF_{\rho}\right] \1_{\{R>\rho \wedge T\}} Y^\rho_T \1_{\{S>\rho\wedge T\}}\right). \nonumber
    \end{align}
    This equality holds $\Q$- and $\widehat{\Q}$-almost surely.
\end{proposition}
\begin{proof}
    Without loss of generality,  assume that $x_0 = 1$. Then, \eqref{E Bayes1} holds $\widehat{\Q}$-almost surely since
    $\widehat{\Q}(S > \rho\wedge T) = 1$ and
    \begin{align*}
        \E^{\widehat{\Q}} \left[\1_A \left(Z \1_{\{R>\tau \wedge T\}}\right) Y^\tau_T\right] &= \E^{\Q} \left[\1_A \left(Z \1_{\{S>\tau\wedge T\}}\right)\right]
            = \E^{\Q} \left[\1_A \E^{\Q}[Z \1_{\{S>\tau \wedge T\}} |\CF_\rho]\right] \\
            &= \E^{\widehat{\Q}} \left[\1_A  \E^{\Q}[Z \1_{\{S>\tau \wedge T\}}|\CF_\rho] \1_{\{R>\rho \wedge T\}} Y^\rho_T \right]
    \end{align*}
    for all $A \in \CF_\rho.$
   Moreoever, \eqref{E Bayes1} holds $\Q$-almost surely since  $\Q(R>\rho \wedge T) = 1$ and
    \begin{align*}
        \E^{\Q} \left[\1_A \left(Z \1_{\{S>\tau \wedge T\}}\right) Y_T^\rho\right] &= \E^{\widehat{\Q}} \left[\1_A (Z \1_{\{R > \tau \wedge T\}}) Y^\rho_T Y^\tau_T\right]\\
            &= \E^{\widehat{\Q}} \left[\1_A \E^{\widehat{\Q}}\left[\left. (Z \1_{\{R > \tau \wedge T\}}) Y^\tau_T \right|\CF_\rho\right] {Y^\rho_T} \right] \\
            &= \E^{\Q} \left[\1_A \E^{\widehat{\Q}}\left[\left.(Z \1_{\{R > \tau \wedge T\}}) Y^\tau_T \right|\CF_\rho\right] \1_{\{S>\rho \wedge T\}}\right]
    \end{align*}
   for all $A \in \CF_\rho.$
\qed\end{proof}

Bayes's formula yields a simple proof of Proposition~\ref{P Equivalence}:
\begin{proof}[of Proposition~\ref{P Equivalence}]
    The statement in (i) is a corollary of Proposition~\ref{P Bayes} if we replace $\tau$ by $\tau \wedge t$ and use
    $Z=N_t^\tau$ and $\rho = \tau \wedge s$ in \eqref{E Bayes1} for all
    $t \in [0,T]$ and $s \in [0,t].$
    
    Assume now that $\{N_t \1_{\{S>t\}}\}\tInd$ is a $\Q$-local martingale on $[0,S)$. Then
    there exists a non-decreasing sequence of  stopping times $\{\tau_i\}_{i \in \N}$ such that
    $\Q(\lim_{i \uparrow \infty} \tau_i = S) = 1$ and that $\{N^{\tau_t} \1_{\{S > \tau_i \wedge t\}}\}\tInd$ is a $\Q$-martingale for all $i \in \N$. 
    Now, (i) implies that $N^{\tau_i} Y^{\tau_i}$ is a $\widehat{\Q}$-martingale.  An application of
    Lemma~\ref{L stoppingtimes} with $\tau := \lim_{i \uparrow \infty} \tau_i$ yields that $N Y$ is a $\widehat{\Q}$-local martingale on $[0,R)$. The reverse direction follows in the same manner. This yields (ii).
    
	Assume next that $\{N_t^{S_i} \1_{\{S>S_i \wedge t\}}\}\tInd$, and, thus, 
	$\{N_t^{S_i^Y} \1_{\{S>S_i^Y \wedge t\}}\}\tInd$ are $\Q$-martingales
	 for all $i \in \N$. Then the statement in (iii) follows from (i) and the fact that 
	$\widehat{\Q}(\lim_{i \uparrow \infty} S^Y_i > T) = 1$ by (i) in Lemma~\ref{L locsequence2}.	
\qed\end{proof}

We conclude this appendix by providing a Girsanov-type result.  Towards this end, let us denote the quadratic
covariation process of two $\Q$-semimartingales  $N^{(1)}$ and $N^{(2)}$ with \cadlag{} paths by ${[ N^{(1)},N^{(2)}]} =
{\{[ N^{(1)},N^{(2)}]_t\}\tInd}.$   If $X$ has \cadlag{} paths, the process $N^{S_i}$ is a $\Q$-semimartingale with \cadlag{} paths, and $[N,X]^{S_i} := [N^{S_i},X]$ has $\Q$-integrable variation for all $i \in \N$, then the quadratic covariation process $[N,X]$ has a compensator ``up to time $S$,'' that is, there exists a process $\langle N,X\rangle = {\{\langle N,X\rangle_t\}\tInd}$ such that $\langle N,X\rangle^{S_i}$ is the compensator of $[N,X]^{S_i}$ for all $i \in \N$; see also Theorem~III.3.11 of Jacod and Shiryaev \cite{JacodS}.  For any \cadlag{} stochastic process $Z = \{Z_t\}\tInd$, we define $Z_{t-} := \lim_{s \uparrow t} Z_s$ for all $t \in (0,T)$ and $Z_{0-} := Z_0$. 

\begin{proposition}[Girsanov-type theorem]  \label{P Girsanov}
	Assume  the setup of Theorem~\ref{T numeraire} and let  $N = \{N_t\}\tInd$ denote a  progressively measurable stochastic process taking values in $[-\infty,\infty]$ such that $N_t = N_t \1_{\{R>t\}}$ for all $t \in [0,T]$ and such that $N^{S_i}$ is a $\Q$-semimartingale with \cadlag{} paths for all $i \in \N$. 
Suppose that $X$ has \cadlag{} paths. 
	We then have the following statements:
    \begin{enumerate}
        \item[(i)] The process $N^{R_i}$ is a $\widehat{\Q}$-semimartingale with \cadlag{} paths for all $i \in \N$.
        \item[(ii)] If $N$ is a $\Q$-local martingale on $[0,S)$  (equivalently, on $[0, R \wedge S)$) and if $[N,X]^{S_i}$ has $\Q$-integrable variation for all $i \in \N$, then
		$\widetilde{N} = \{\widetilde{N} _t\}\tInd$, defined by 
    \begin{align*} 
        \widetilde{N} _t := N_t - \int_0^{t} Y_{s-} \dd \langle N,X \rangle_s
    \end{align*}
	for all $t \in [0,T]$,
	 is a $\widehat{\Q}$-local martingale on $[0,R)$  (equivalently, on the interval $[0, R \wedge S)$).
        \item[(iii)] If $N$ is a $\Q$-local martingale $[0,S)$  (equivalently, on $[0, R \wedge S)$) and if we have $\Q(S>S_i \wedge T) = 1$ for all $i \in \N$, then
		$\widehat{N} = \{\widehat{N} _t\}\tInd$, defined by 
    \begin{align*}  
        \widehat{N} _t := N_t - \int_0^{t \wedge S} Y_s \dd [ N,X ]_s
    \end{align*}
	for all $t \in [0,T]$,
	 is a $\widehat{\Q}$-local martingale on $[0,R)$  (equivalently, on the interval $[0, R \wedge S)$).
        \end{enumerate}
\end{proposition}
\begin{proof}
	The proof is based on a simple localization argument. Observe that $\dd \widehat{\Q}|_{\CF_{R_i} \cap \CF_{R-}} = X^{R_i}_T \dd {\Q}|_{\CF_{R_i} \cap \CF_{R-}}$; to wit, $\widehat{\Q}$ is absolutely continuous with respect to ${\Q}$ on $\CF_{R_i} \cap \CF_{R-}$ for all $i \in \N$.
	Thus, (i) corresponds directly to Theorem~III.3.13 in Jacod and Shiryaev  \cite{JacodS}.  By Theorems~III.3.11 in Jacod and Shiryaev  \cite{JacodS}, the process $\widetilde{N}^{R_i}$ is a $\widehat{\Q}$-local martingale; thus $\widetilde{N}$ is a $\widehat{\Q}$-local martingale on $[0,R_i)$ for all $i \in \N$. Lemma~\ref{L locsequence} then yields (ii). Similar reasoning yields that $\widehat{N}$ is a $\widehat{\Q}$-local martingale on $[0,R)$  by applying Theorem~3 in Lenglart \cite{Lenglart_1977} after observing that the proof therein also works for probability spaces that do not satisfy the usual assumptions.
%
\qed\end{proof}

\begin{remark}[Lack of martingale property in Proposition~\ref{P Girsanov}]
    One might wonder whether (ii) or (iii) of Proposition~\ref{P Girsanov} can be strengthened by replacing
    each ``local martingale'' by ``martingale.''  Example~\ref{example LackEquiv} illustrates that such a statement would be false, even in the case of $X$ being a
    strictly positive, true $\Q$-martingale. To see this,
    replace $\widehat{\Q}$ by $\Q^{Z}$ and the processes $N$ by $X$ and
    $X$ by $Z$ in Proposition~\ref{P Girsanov}. Then
    $N$ is a true $\Q$-martingale but $\widetilde{N} = \widehat{N} = N$ is
    only a strict $\Q^Z$-local martingale.
    \qed
\end{remark}

\section{Proof of Lemma~\ref{L consistency}}  \label{A proof 4}
In this appendix, we will provide the proof of Lemma~\ref{L consistency}:
\begin{proof}[of Lemma~\ref{L consistency}]
	The fact that (ii) implies (i) follows directly from \eqref{E Qe} and \eqref{E QdQe} with $Z = \1_A$ and $\tau = R_i \wedge S_j$ for all $A \in \CF_{R_i \wedge S_j}$ and $i,j \in \N$
	since $\Qe(R>R_i \wedge T) = 1 = \Qd(S>S_j \wedge T)$. 

For the reverse direction, fix a stopping time $\tau$ and note that it is sufficient to show \eqref{E Qe0} for such events $A \in \CF_{\tau \wedge T}$ that satisfy $$A = A \cap \{R > \tau \wedge T\} \cap \{S > \tau \wedge T\}$$ since $\Qe(S \leq T) = 0$ as $Y$ is a $\Qe$-local martingale and, thus, cannot explode. Therefore, we may assume, without loss of generality, that $A \in \CF_{(R \wedge S)-}$.
Let $\widehat{\Qd}$ denote the unique probability measure on $(\Omega, \CF_{R-})$ that was constructed in Theorem~\ref{T numeraire} with $\Q$ replaced by $\Qd$. We need to show the identity $\widehat{\Qd}|_{\CF_{(R \wedge S)-}} = {\Qe}|_{\CF_{(R \wedge S)-}}$. 

Since $\bigcup_{i,j \in \N}  \CF_{(R_i \wedge S_j)-}$ is a $\pi$-system that generates $\CF_{(R \wedge S)-}$ it is sufficient to show that $\widehat{\Qd}|_{\CF_{(R_i \wedge S_j})-} = {\Qe}|_{\CF_{(R_i \wedge S_j)-}}$ for 
all $i,j \in \N$.  Next, fix $i, j \in \N$ and note that, by (i), $\widehat{\Qd}$ and $\Qe$ are equivalent on $\CF_{(R_i \wedge S_j)-}$. Therefore, the $\Qe$-martingale $Z = \{Z_t\}\tInd$ with $Z_t := \dd \widehat{\Qd}/\dd \Qe|_{\CF_t \cap \CF_{(R_i \wedge S_j)-}}$ for all $t \in [0,T]$ is well-defined. We need to show that $Z_T = 1$. Observe that the measure $\widetilde{\Qe}$, defined by $\dd \widetilde{\Qe} / \dd \Qe = Z_T$ is also equivalent to $\Prob^{\eu}$
and the processes $S^{\eu}$ are $\widetilde{\Qe}$-local martingales; see also Proposition~\ref{P Equivalence}. Since $\Qe$ was assumed to be unique among these measures, we may conclude.
\qed\end{proof}

\bibliographystyle{spmpsci}
\bibliography{aa_bib}{}
\end{document}